\documentclass[12pt,preprint]{aastex}

\newcommand{\bpic}{$\beta$~Pic }

\newcommand{\change}{}

\slugcomment{Astrophysical Journal, in press}

\shorttitle{The Disk and the Planet of $\beta$~Pictoris}
\shortauthors{Apai et al.}

\begin{document}


\title{The Inner Disk Structure, Disk-Planet Interactions, and Temporal Evolution in the $\beta$~Pictoris System: A Two-Epoch HST/STIS Coronagraphic Study}


\author{D\'aniel Apai\altaffilmark{1} and Glenn Schneider}
\affil{Department of Astronomy and Steward Observatory, The University of Arizona, Tucson, AZ 85721}
\email{apai@arizona.edu}

\author{Carol A.  Grady\altaffilmark{2,3}}
\affil{Eureka Scientific, 2452 Delmer, Suite 100, Oakland CA 96002, USA}

\author{Mark C. Wyatt}
\affil{Institute of Astronomy, University of Cambridge, Madingley Road, Cambridge CB3 0HA, UK}

\author{Anne-Marie Lagrange\altaffilmark{4}}
\affil{Universit\'e Grenoble Alpes, IPAG, 38000, Grenoble, France; }

\author{Marc J. Kuchner and Christopher J. Stark}
\affil{NASA Goddard Space Flight Center, Exoplanets and Stellar Astrophysics Laboratory, Code 667, Greenbelt, MD 20771, USA}


\author{Stephen H. Lubow}
\affil{Space Telescope Science Institute, 3700 San Martin Drive, Baltimore, MD 21218, USA}


\altaffiltext{1}{Department of Planetary Sciences and Lunar and Planetary Laboratory, The University of Arizona, Tucson AZ 85718}
\altaffiltext{2}{ExoPlanets and Stellar Astrophysics Laboratory, Code 667, Goddard Space Flight Center, Greenbelt, MD 20771, USA} 
\altaffiltext{3}{Goddard Center for Astrobiology, Goddard Space Flight Center, Greenbelt, MD 20771, USA}
\altaffiltext{4}{CNRS, IPAG, 38000, Grenoble, France}


\begin{abstract}
We present deep HST/STIS coronagraphic images of the \bpic debris disk obtained at two epochs separated by 15 years. The new images and the re-reduction of the 1997 data provide the most sensitive and detailed views of the disk at optical wavelengths as well as the yet smallest inner working angle optical coronagraphic image of the disk. Our observations characterize the large-scale and inner-disk asymmetries and we identify multiple breaks in the disk radial surface brightness profile. We study in detail the radial and vertical disk structure and  show that the disk is warped. We explore the disk at the location of the \bpic b super-jupiter and find that the disk surface brightness slope is continuous between 0\farcs5 and 2\farcs0, arguing for no change at the separations where \bpic b orbits.

The two epoch images constrain the disk's surface brightness evolution on orbital and radiation pressure blow-out timescales. We place an upper limit of 3\% on the disk surface brightness change between 3-5", including the locations of the disk warp, and the CO and dust clumps. 

We discuss the new observations in the context of high-resolution multi-wavelength images and divide the disk asymmetries in two groups: axisymmetric and non-axisymmetric. The axisymmetric structures (warp, large-scale butterfly, etc.) are consistent with disk structure models that include interactions of a planetesimal belt and a non-coplanar giant planet. The non-axisymmetric features, however, require a different explanation.
\end{abstract}


\section{Introduction}

Debris disks represent the late stage of planet formation when the primordial disk has dissipated and the accretion of giant planets has concluded, but rocky planets may still continue to accrete from a disk of icy/rocky minor bodies. Collisional cascades of the small bodies generate copious amounts of fine dust (e.g., \citealt[][]{Gaspar2012}), often detectable in thermal emission and sometimes in scattered light.  With typical ages between ten to a few hundred million
years, bright debris disks provide insights into the properties and evolution of young planetary systems.
The distribution of fine dust in these systems can reveal the structure of the underlying planetary systems (e.g. \citealt[][]{StarkKuchner2008,Apai2008,Roberge2009,Su2013}) 
and allows statistical studies of the planetary system properties, dynamics, and collisional history (for reviews see \citealt[][]{Meyer2007,Wyatt2008,MatthewsPPVI}). 
In addition, debris disks likely pose a critical limitation on future missions to directly image terrestrial exoplanets \citep[][]{Guyon2006,Millan-Gabet2011}.

However, debris disks around stars known to host exoplanets remain very rare, making it difficult to establish a quantitative link between disk structures and planets. \bpic is a unique system hosting the best-studied debris disk and in it a directly imaged super-jupiter. This system is also nearby (19.44$\pm$0.05~pc, \citealt[][]{vanLeeuwen2007}), allowing high physical resolution with current instrumentation.

The \bpic disk is massive and exceptionally bright ($F_{disk}/F_{\star}$ = $\sim$2.5$\times10^{-3}$, \citealt[][]{Lagrange2000}),
with a diameter exceeding 400 au (e.g. \citealt[][]{SmithTerrile1984}). 
The disk has been imaged with multiple instruments, but
the most spectacular and most detailed images have been obtained with
HST/STIS and HST/ACS/HRC (\citealt[][]{Heap2000,Golimowski2006}).
Perhaps the most perplexing structure in \bpic is a warp {\em or} secondary disk inclined $\simeq5^\circ$
to the main disk.
The STIS coronagraphic images have been interpreted as a warped disk \citep[][]{Heap2000}, while
\citet[][]{Golimowski2006} have used deconvolved HST/ACS images to argue for the presence of a primary and a smaller, fainter, and inclined secondary disk. In the deconvolved ACS images the two wings of the main disk show prominent NW-SE asymmetry and at least three
breaks between 40 and 250 au in its radial surface brightness distribution, indicative of
changes in the disk structure or in dust grain properties. These breaks in surface brightness
distribution may emerge from dust belts and/or planetesimals belts \citep[e.g.][]{Wilner2011}.
The NW and SW wings of the disk also have different optical colors; in addition, while the
NW wing is linear, the SW wing is bowed. 

 In the HST/ACS images the two wings of the secondary disk follow similar radial brightness profiles, suggesting 
identical radial density distributions and dust grain populations  \citep[][]{Golimowski2006}.
Surprisingly, however, these two linear wings are not collinear.
In summary, the scattered-light \bpic disk has a complex structure with evidence for different grain
populations and radial density profiles, as well as breaks, discontinuities or asymmetries in
the central region of the disk. Planetary perturbations have been invoked to explain the large-scale asymmetries in the main disk or the secondary
disk/warp \citep[e.g.][]{Mouillet1997, Augereau2001,Golimowski2006}.


Recently, \citet[][]{Lagrange2009} and \citet[][]{Lagrange2010} announced the discovery and
confirmation of a giant planet orbiting $\beta$~Pictoris. The planet has an estimated mass of 8--10 M$_{\mathrm J}$ 
and an orbital semi-major axis of about 8 au \citep[][]{Chauvin2012}. Since the discovery of the planet its motion has
been closely monitored and its orbit has been gradually refined \citep[e.g.][]{Quanz2010,Bonnefoy2011,Chauvin2012, Males2014,Nielsen2014,Macintosh2014}.
The planet's orbit was found to be slightly inclined ($\sim1^\circ$) with respect to the main disk and to have a relatively low eccentricity (e$<$0.1, \citealt[][]{Lagrange2012,Macintosh2014}). 

Recent observations with Herschel \citep[][]{Vandenbussche2010} and SMA \citep[][]{Wilner2011} 
sampled the spectral energy distribution of \bpic at far-infrared and millimeter wavelengths and provided resolved
or marginally resolved images of the disk. Most interestingly, \citet[][]{Wilner2011} identified
two millimeter clumps at radii $\sim$3\farcs5 and argued that these indicate a population of
planetesimals in a ring at $r=94\pm8$~au -- inclined with respect to the midplane --  whose collisions serve as a source of dust grains.
A few months ago \citet[][]{Dent2014} published ALMA high-resolution (12 au) and high sensitivity 870~$\mu$m continuum and $^{12}$CO 3-2 transition observations of the \bpic disk. The continuum images revealed that the southwestern disk is brighter than the northeastern one and shows a peak at 60 au (3\farcs08) with a tail extending beyond 4\farcs0. The CO distribution also shows excess emission in the SW side, but it peaks at a larger separation (85 au or 4\farcs4). Unlike the continuum emission, the CO clump detected by ALMA peaks above the disk midplane ($\sim$5 au). The velocity-based de-projection of the CO data argues for, but does not prove, a broad CO clump at the Earth-facing SW quarter of the disk at with an orbital radius of 85~au and a CO tail extending toward the NE side of the disk \citep[][]{Dent2014}. The dust mass is estimated to be $4.7\pm0.5 \times 10^{23}$~kg (6.4 M$_{Moon}$, while the CO mass is approximately $1.7 \times 10^{20}$~kg (0.0023 M$_{Moon}$). The presence of such large amounts of CO gas is surprising, given its short dissociation lifetime ($\sim$120~yrs) and poses a problem similar to that seen in an another old debris disk (HD 21997, \citealt[][]{Kospal2013}).

\citet[][]{Dent2014} identify two possible scenarios to explain the morphology of the CO and continuum observations: In the first, the clumps emerge from collisions of planetesimals trapped in 2:1 and 3:2 mean motion resonance with an outward migrating $>$10~M$_{Earth}$ planet \citep[][]{Wyatt2003}. In the second, a single, massive, and recent ($\sim$0.5~Myr) collision with an approximately Mars-sized parent body injects dust and CO gas into the system. 

In a 2014 VLT adaptive optics study \citet[][]{Milli2014} imaged the disk between 0\farcs4 and 3\farcs8 in L$^\prime$. They report an overall bowed disk structure, particularly prominent in the inner disk ($<$1" from the star). Based on an {\change anisotropic} scattering model they explain the key features in the disk surface brightness distribution (bowed structure, surface brightness profile, and the warp) by a two-component disk: an outer (main) disk extending outward from an inner radius ($r_{warp}$), and an inner warped disk inclined at 4$^\circ$ with respect to the main disk. In their model, the entire disk is inclined toward the observer at an inclination angle 86$^\circ$. 
 
In this study we explored two questions important to understanding
the $\beta$~Pictoris system: {\em i) What is the structure of the inner disk?}, and {\em ii) Are there planetesimal groups in orbits
resonant or co-moving with $\beta$~Pic~b?} We addressed these questions by obtaining high-quality coronagraphic images
of the system using the Hubble Space Telescope. The new optical disk images have the smallest yet inner working angles ($\sim0\farcs35$), allowing
detailed studies of the inner disk in the absence of image artifacts at small stellocentric distances previously probed. We also re-reduced data from 1997, obtained in a similar mode but with a larger inner
working angle, (now) with PSF-template subtracted coronagraphy. By comparing the disk structure in the two epochs we searched for temporal evolution in the inner disk over
timescales comparable to the orbital period of the planet and similar to radiation pressure timescales.

The paper is organized as follows. We first review the observations, data reduction and basic
data analysis (Section~\ref{Sect:Observations}). Section~\ref{DiskAngle} describes the disk's orientation and overall shape, Section~\ref{DiskSurfaceBrightnessProfiles} 
discusses the surface brightness profiles along various directions and the large-scale asymmetry of the disk, Section~\ref{VerticalStructure} explores the vertical structure of the disk and the warp through various methods, Section~\ref{S:Orbit} compares the orbit of the planet to the inner disk structure  and explores temporal evolution of the scattered light disk. Section~\ref{S:DiskStructure} provides a multi-wavelength view of the disk. Section~\ref{Sect:Models} discusses our observational results in the context of disk--planet interaction models. Finally, Section~\ref{Summary} summarizes the key findings of our work.


\section{Observations and Data Reduction}
\label{Sect:Observations}

In this paper we present a re-reduction of a 1997 HST/STIS (Space Telescope Imaging Spectrograph, \citealt[][]{Woodgate1998}) archival dataset  and also introduce a new set of HST/STIS coronagraphic observations we obtained in 2012. STIS coronagraphy utilizes an image plane mask including two orthogonal wedge occulters that is inserted in the STIS focal plane. STIS cannot use filters in its coronagraphic (50CORON) mode; thus, the images described below are
unfiltered data whose wavelength coverage is only set by the spectral response of the STIS CCD detector ($\sim$200-1,050 nm with $\lambda_{pivot}$=575.2 nm and FWHM=433 nm and a spectrally flat source). Based on the STIS Exposure Time Calculator 22.1.1 and assuming an input stellar spectrum of A5V (corresponding to \bpic) we find that the effective wavelength (the wavelength at which incoming photons generate the most electrons) is 571.3~nm.
 The STIS CCD detector is a 1024$\times$1024 pixel UV-enhanced chip with an image pixel scale of 0\farcs05077 per pixel and at these wavelengths a resolution element is $\sim$60~mas.

\subsection{The 1997 STIS Observations}

The 1997 observations and their analysis are described in detail in \citet[][]{Heap2000} and we only give a brief  summary of the data we have used here.  Table~\ref{T:observations} summarizes the observations.The first epoch observations were carried out in program GO-7125 (PI: Heap) on Sep 16-17, 1997 with STIS occulting wedge B. \bpic was observed in three non-contiguous orbits (program visits \#4-6, archived raw dataset IDs: O4204, 05, 06). Prior data from visits 1--3 using wedge A failed or were degraded and ill-suited for an investigation.
In the first wedge B orbit the disk was placed halfway between the telescope's diffraction spikes  and (nearly) orthogonal to the occulting wedge, while the second and third orbits were executed with spacecraft off-rolled by $-12^\circ$ and $+14^\circ$ relative to the first orbit. During the observations \bpic was occulted by WedgeB2.0 and WedgeB1.0. At each location
a series of images were taken with 5~s and 3~s integration times, respectively. Here we only re-reduced the WedgeB1.0 images, for lack of a suitable Wedge B2.0 reference PSF.

\subsection{The 2012 STIS Observations}

Our new STIS observations were carried out on March 6, 2012 and used a refined observing strategy that  resulted in higher signal-to-noise, weaker PSF residuals, and even smaller inner working angle. Our observing strategy followed \citet[][]{Schneider2014}, using multiple roll angles, coronographic wedge positions, and contemporaneous observations of a color- and brightness-matched reference star to minimize PSF residuals.

The data were acquired in program GO-12551 (PI: Apai). Table~\ref{T:observations} gives a summary of the observations. In this  program using three contiguous orbits we targeted \bpic in the first and third orbits and the PSF reference star ($\alpha$~Pic) in the second orbit.
The PSF star ($\alpha$~Pic, V=3.3, B$-$V= 0.18) was selected to closely match the celestial position, apparent magnitude, and color of \bpic (V=3.86, B$-$V=0.17).  We executed spacecraft rolls between the three orbits to recover (in combination) azimuthal sectors in the celestial frame otherwise unimaged due to the imposition of the coronagraphic wedge or degraded by the telescope diffraction spikes, and to further reduce the rotationally non-invarient components of the PSF-subtraction residuals.
In each of the three orbits we used a combination of short, medium, and long integrations (13.2\,s, 48\,s, 240\,s) at 0\farcs6 and 1\farcs0 wedge width locations; while the
0\farcs6 location was observed in WedgeA, the 1\farcs0 wedge location was observed both at the A- and B coronagraphic wedges of STIS.
The inner working angle (IWA) equals the half-width of the occulting wedge at the stellar position: WEDGE(A or B)0.6 = 0\farcs3 IWA and WEDGE{A|B}1.0=0\farcs5 IWA. 
{\change Due to the very high contrast in the image between the star, inner disk, and the outer disk no single image has suitable dynamic range to capture the scattered light surface brightness levels at all radii; therefore, we used short integrations to probe the stellar PSF and the inner disk, and longer integration times to probe the outer disk (while saturating the inner disk). The combination of the non-saturated pixels in images with different integration times allowed us to significantly increase the effective dynamic range of our final images and to probe the disk with high signal-to-noise over a surface brightness range of more than $10^5$. For more details on the image combination we refer the readers to \citep[][]{Schneider2014}. }

\begin{table}
\begin{center}
\caption{Summary of the STIS observations used in this work.\label{T:observations}}
\begin{tabular}{lccccccc}
\tableline\tableline
Prog. & Visit \# & Target & Date & Int. Times [s] & ORIENT$^a$ & Aperture  \\
\tableline
7125 &  04  & $\beta$~Pic  &  1997-09-17   & 16$\times$3  & 210.07         & WEDGEB1.0        \\
7125 &  05 & $\beta$~Pic  &  1997-09-17  & 16$\times$3   & 224.07     & WEDGEB1.0            \\
7125 &  06 & $\beta$~Pic  &  1997-09-17  & 16$\times$3 & 198.07       & WEDGEB1.0                            \\
\hline
12551 & 01 & $\beta$~Pic  & 2012-03-6 &  11$\times$1.2                &    239.15     & WEDGEA0.6  \\
12551 & 01 & $\beta$~Pic  &  2012-03-6  & 16$\times$3.0, 4$\times$60.0 &    239.15      & WEDGEA1.0 \\
12551 & 01 & $\beta$~Pic  &  2012-03-6  &  4$\times$60.0, 16$\times$3.0 &    239.15      & WEDGEB1.0       \\
\hline
12551 & 02 & $\alpha$~Pic  &  2012-03-6& 11$\times$0.7    &   245.96    & WEDGEA0.6                   \\
12551 & 02 & $\alpha$~Pic  &   2012-03-6& 4$\times$36.0, 16$\times$1.9   &   245.96      & WEDGEA1.0                      \\
12551 & 02 & $\alpha$~Pic  &  2012-03-6& 16$\times$1.9, 4$\times$36.0&   245.96      & WEDGEB1.0    \\
12551 & 02 & $\alpha$~Pic  &  2012-03-6& 11$\times$0.7, 2$\times$(3$\times$0.7)$^b$ &   245.96      & WEDGEB0.6          \\
\hline
12551 & 03 & $\beta$~Pic  &  2012-03-6 &  11$\times$1.2      &    272.62     & WEDGEA0.6                    \\
12551 & 03 & $\beta$~Pic  &  2012-03-6 & 16$\times$3.0, 4$\times$60.0     &   272.62      & WEDGEA1.0              \\
12551 & 03 & $\beta$~Pic  & 2012-03-6& 4$\times$60, 16$\times$3.0     &    272.62    & WEDGEB1.0        \\
\tableline
\end{tabular}
{$^a$}{ORIENTAT: position angle of the image +Y axis measured eastward from celestial north.}
{$^b$}{Additional two sets of three exposures each with ${1 \over 4}$ pixel in Y offsets for calibration purposes.}
\end{center}
\end{table}

\begin{figure}
\epsscale{1.20}
\plottwo{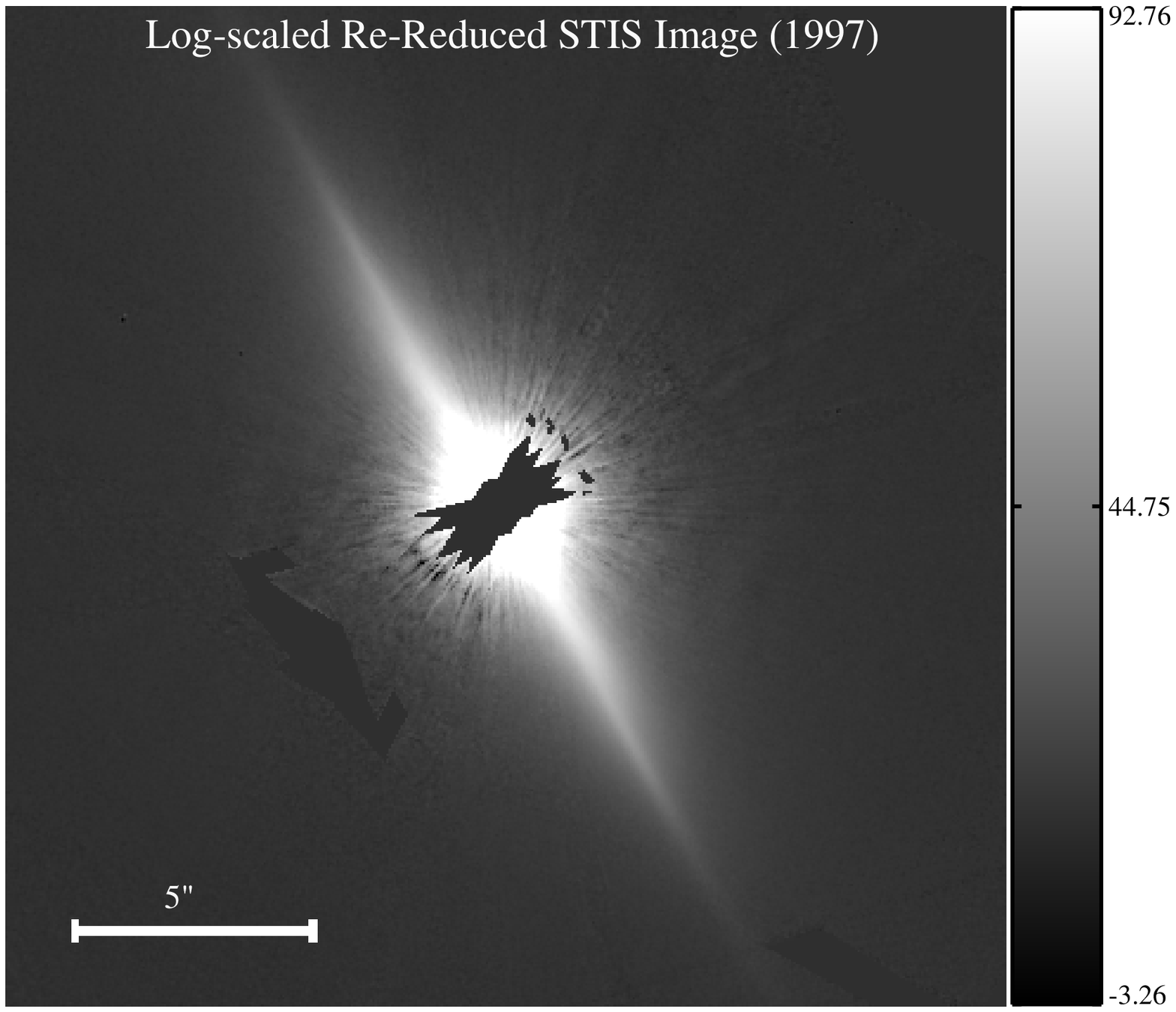}{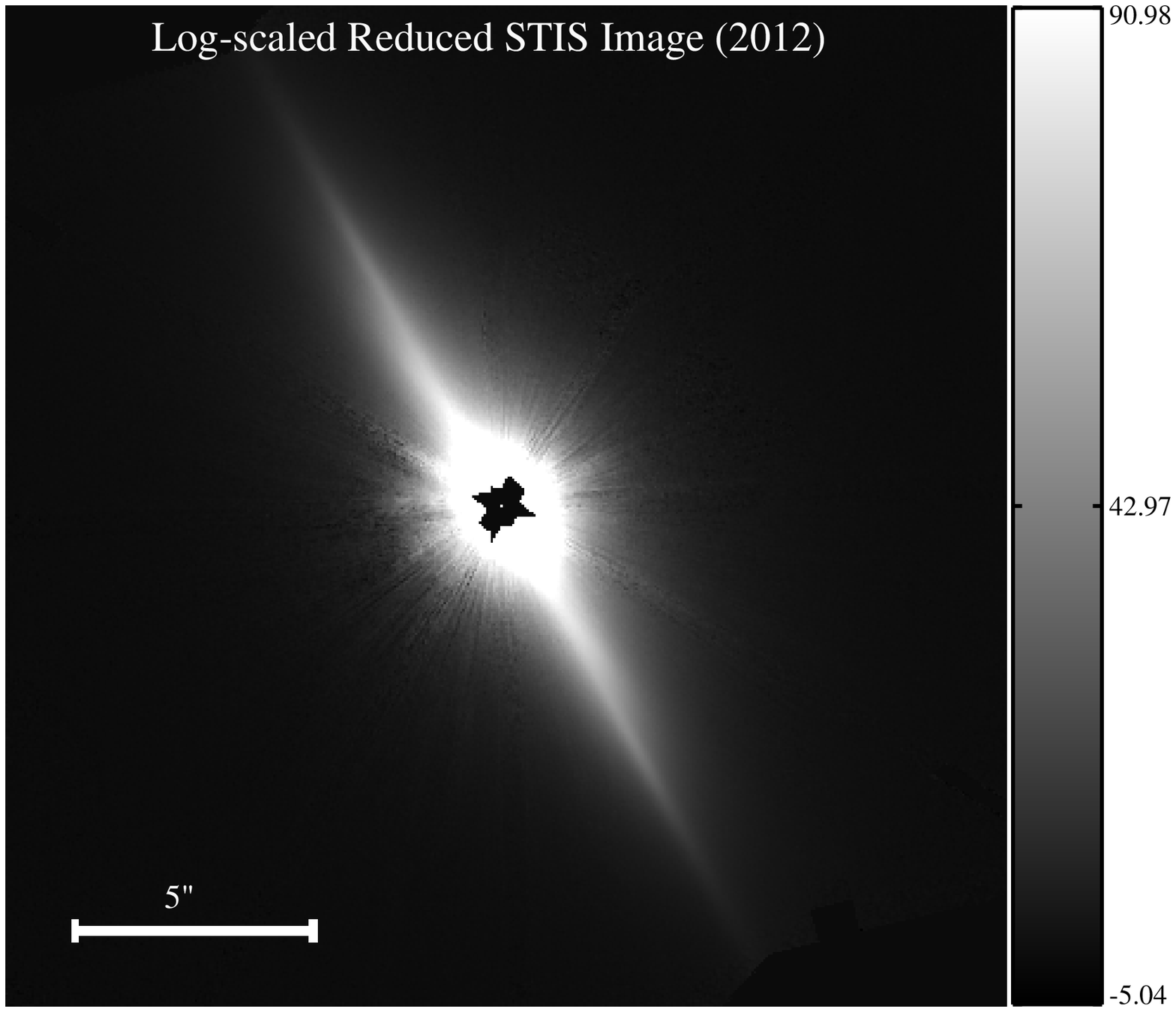}
\caption{Direct comparison of the two STIS data sets on the $\beta$~Pictoris disk. {\em Left:} Our re-reduction of the 
STIS image taken in GO-7125 in 1997. {\em Right:} New STIS images taken in our program GO-12551 provide
higher signal-to-noise and better PSF-subtraction, as well as smaller inner working angle.  The two images are shown
here in a logarithmic stretch, with instrumental brightness units of counts per second per pixel; North is up. \label{Fig-Complete}}
\end{figure}

\subsection{Coronagraphic Data Reduction}

Our data reduction follows that of \citet[][]{Schneider2009}. Here we provide a brief summary of the key steps
and differences from that dataset and note that the GO 7125 and GJ12551 images were reduced separately.

First, the {\tt RAW} files were reprocessed with the STIS basic calibration package {\tt calstis} using the
most up-to-date dark and bias reference files closely contemporaneous with the observations. Each instrumentally calibrated image was then manually inspected for anomalies, but
none were found. Within each visit groups of exposures with the same setup (target, orientation, pointing, aperture, and exposure time) were median combined. Next, for each median-combined image we used small sub-arrays located far from the target to estimate
the combination of the sky and instrumental background, which were then subtracted from the images. 
By dividing each background-subtracted median combined image with the exposure of the individual images we calculated the count rates per pixel. We next
located the positions of the target (occulted star) by identifying the intersection of two lines fit to the diffraction
spikes (see \citealt[][]{Schneider2014} for details of the process).

The above steps have been repeated on every frame taken on \bpic and the PSF reference star. 
In the next steps from each \bpic image we subtracted a matching PSF template. Because in the GO~7125 no PSF template star observations were taken, from these images we subtracted the WedgeB PSF template derived from the 2012 epoch GO~12551 observations. The precise alignments of the target and PSF images are critical and we used a two-step alignment procedure. First, the target and template images have been co-aligned using fractional pixel offsets to the initial centers measured from the diffraction spike-fitting procedure {\change (\citealt[][]{Schneider2009}, \S 4.2)}. 
Then we refined the alignments between the target and PSF images by iteratively minimizing the diffraction spike residuals 
in the PSF-subtracted images where the disk flux did not dominate by varying three free parameters ($\Delta~X$ and $\Delta~Y$, and intensity scaling). We iterated all three parameters until convergence was reached in minimizing the subtraction residuals.  This solution was also verified by visual inspection. This two-step alignment procedure was repeated for every target--PSF image pair.

After precise target and template alignment and PSF subtraction the images were rotated to a common "north up" orientation about the occulted star; image edge clipping was avoided by embedding the images in a larger, zero--padded image. For each image we created a binary "bad" data mask that flagged all pixels that were saturated, corrupted by diffraction spikes, covered or affected by the wedges, or otherwise significantly degraded. For the short Wedge0.6A/B images this mask also included pixels at large stellocentric angles where the signal-to-noise was low. Finally, the rotated, masked images were median combined to create the final analysis-quality images from the GO~7125 and GO~12551 data sets.

Our final images reveal the complex debris disk around \bpic  between 0\farcs35 to about 13", 
providing an image with the highest quality yet and also the smallest working angle (Fig.~\ref{Fig-Complete}) at visible wavelengths.

{\change Figure~\ref{Fig-SNRMap} shows our estimated signal-to-noise maps of the re-reduced 1997 and the new 2012 images. The maps were calculated by dividing the instrumental counts by the noise in the images, both averaged over 3$\times$3 STIS pixels. The noise for each pixel position was estimated as the standard deviation of the counts in the given pixel in each of the frames that covered that pixel. Note, that the reliability of the noise estimate for any given pixel is sensitive to the number of valid pixels that covered that location.}

\begin{figure}
\epsscale{1.0}
\plotone{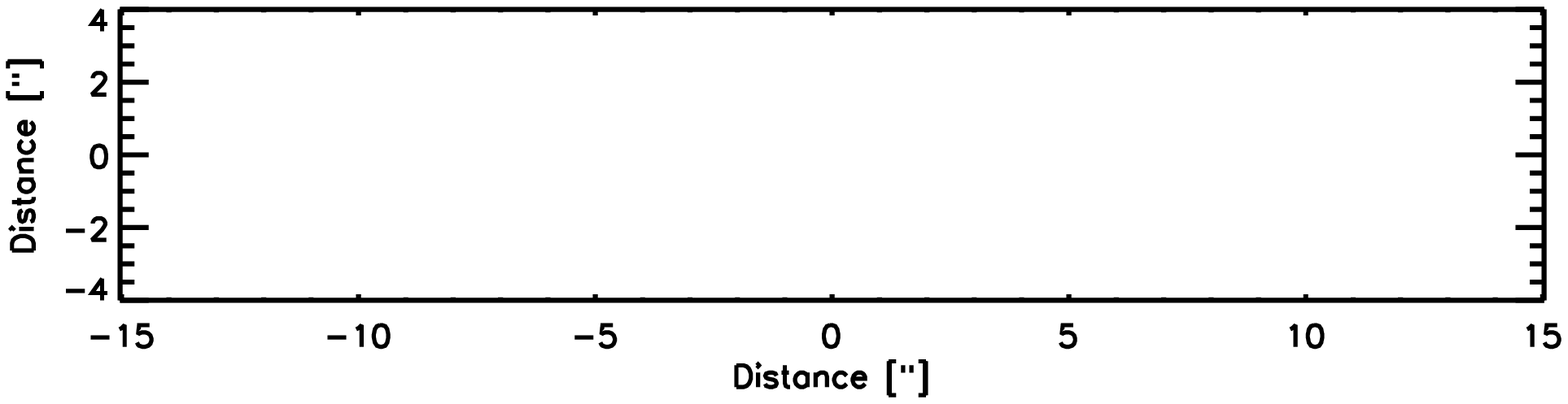}
\caption{Direct comparison of the two disk images obtained from the re-reduction of the 1997 dataset and from
the new 2012 STIS images. While the 1997 data has a slightly larger effective field of view (not shown here), importantly the 2012 data have smaller inner working angle and higher signal-to-noise. These images have been multiplied by an $r^{1.8}$ function to enhance the fainter 
outer disk's visibility while simultaneously showing the inner disk. The image has been rotated counter-clockwise by 60.9$^\circ$; the NE side of the disk is on the left.\label{Fig-BigDisk}}
\end{figure}

\begin{figure}
\epsscale{1.0}
\plotone{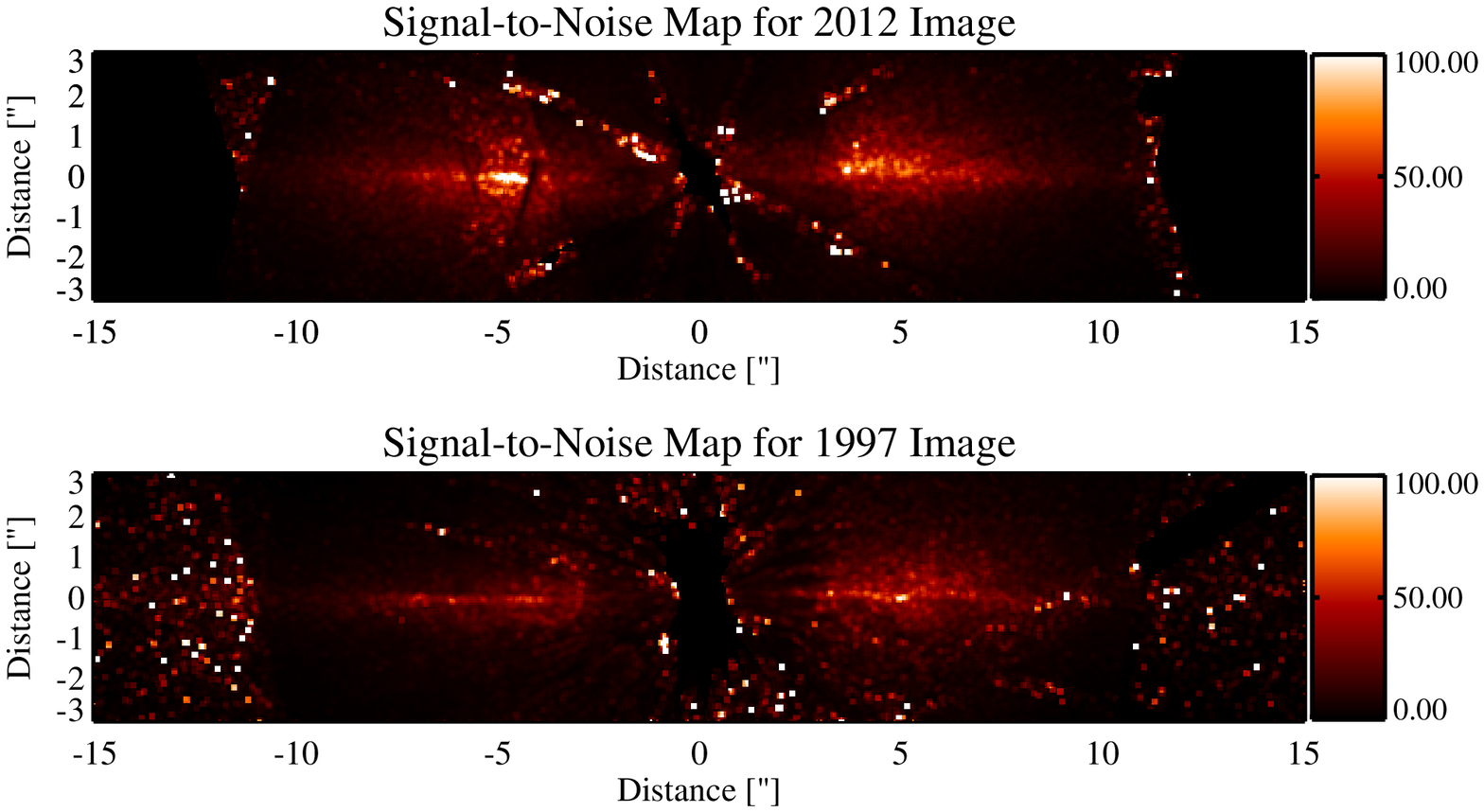}
\caption{Signal-to-noise ratio maps for the combined 2012 and 1997 images. The 2012 images provide higher signal-to-noise ratio across the image and, in particular, at small inner working angles close to the star. The images have been rotated counter-clockwise by 60.9$^\circ$; the NE side of the disk is on the left.\label{Fig-SNRMap}}
\end{figure}

To demonstrate the improvement in the image quality with the GO~12551 data to the re-reduced GO-7125 data we note that the signal-to-noise integrated over 2$\times$2 pixel area in the 1997 images is 20, while it is 40 in the 2012 images (in the disk midplane at separation 35 pixels or 1\farcs77); in the better-sampled closer in regions the improvement is even more significant: the 2012 images reach a signal-to-noise of 120, while the 1997 images have a signal-to-noise of 12 (at 21 pixels or 1\farcs06) after our reprocessing.
In Figure~\ref{Fig-BigDisk} we plot the two images with the main disk axis aligned horizontally (assuming a position angle P.A.=29$^\circ$.1)  and by applying a radial brightness scaling ($\times r^{1.8}$) that allows easy comparison of the inner and outer disk structures. Here the 1.8 exponent is not physically motivated but was chosen for visual illustration in collapsing the radial dynamic display range.


\section{Disk Orientation and Surface Brightness Profile}

\subsection{Disk Position Angle}

\label{DiskAngle}

We determined the main disk's position angle by measuring the directions of vectors pointing from the star's position to the disk isophotically-determined mid-plane at {\change stellocentric} separations of {\change 11.2"} in both the northeastern and southwestern wings {\change for both the 1997 and the 2012 images}. 
Because the disk is not completely straight we did not attempt an algorithmic determination of the disk angle but instead opted for the manual identification of the brightest pixel at the given radii and calculated the slope of the line connecting these points. {\change With this procedure we found a disk position angle of $29.17^\circ$ for the 1997 image and a disk position angle of $29.05^\circ$ for the 2012 image, each with an estimated uncertainty of $\pm0.1^\circ$. This uncertainty corresponds to 1 pixel uncertainty in the position of the disk midplane over the full visible disk in our images, which is realistic given our manual fitting procedure. However, this uncertainty does not include the uncertainty in the spacecraft roll angle, which is typically less than $0.1^\circ$ \citep[][]{Kalas2013}. 
To estimate the uncertainties in the absolute spacecraft roll angle between the two epochs we searched for the minimum of the difference between the x-y aligned 1997 and 2012 images on a finely sampled grid of rotation (0.05$^\circ$/grid point). We found that a 0.0$^\circ$ relative rotation produces the smallest difference, i.e. the HST astrometric reference frame is highly reproducible  between the two images, even though different guide stars were used in the two epochs at different spacecraft orientation angles. This finding reinforces our conclusion that the uncertainty of our disk angle measurement is dominated by the uncertainty in the disk midplane {\em determination}, and not differences in the spacecraft's astrometric reference frame. 
Therefore, we adopt the average of the two position angles as the disk position angle, i.e. we will use 29.1$\pm0.1^\circ$ (counter-clockwise from north).}

\begin{figure}
\epsscale{1.0}
\plotone{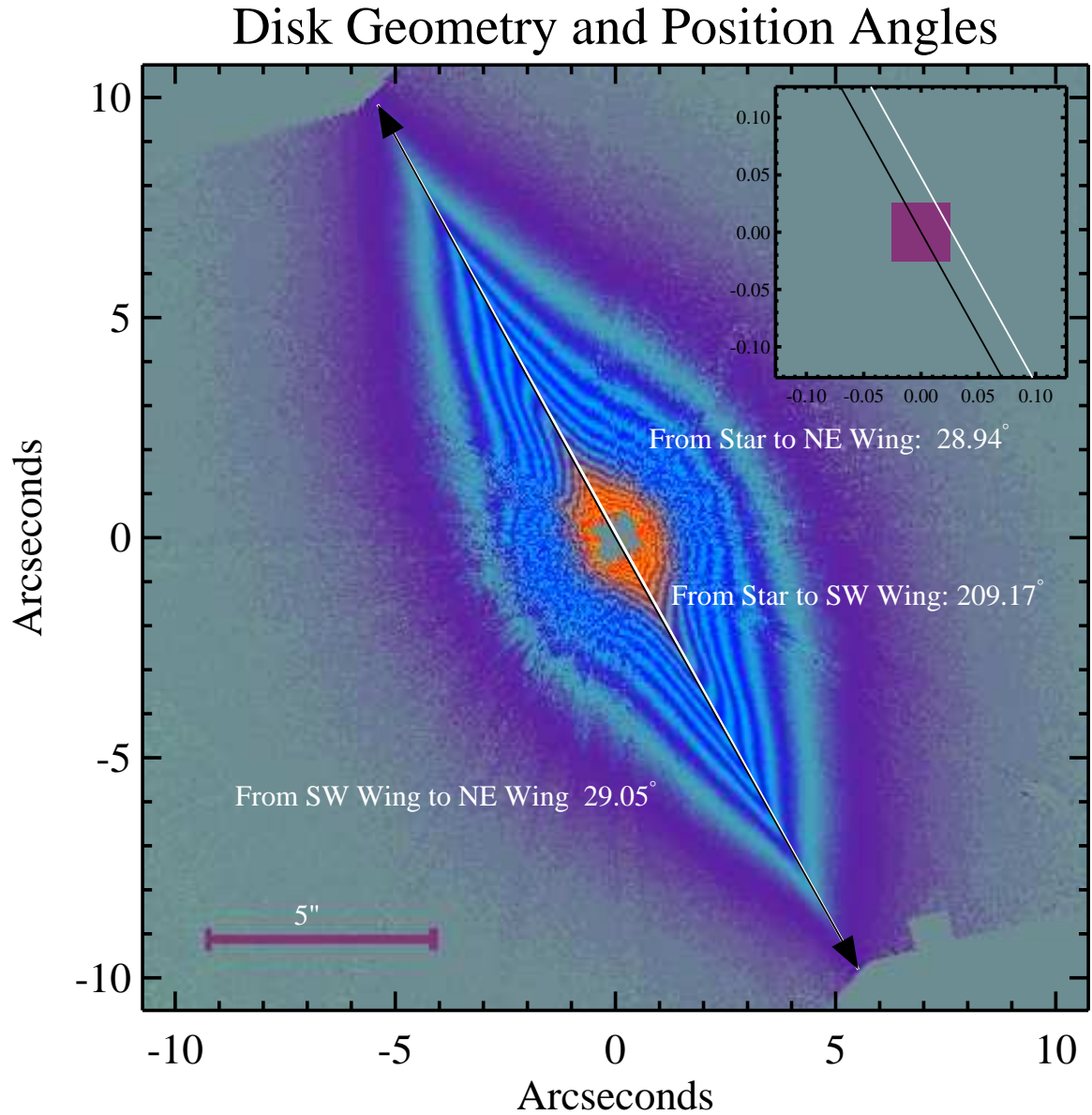}
\caption{{\change Our 2012 image and the illustration of the disk angle determination. The disk position angle was determined by fitting a line (white line) on the brightest points in the disk (disk spine) 11" on either sides of the disk. This line does not precisely intersect the star itself (see inset), demonstrating an overall bow in the disk. The black lines connect the {\em star} to the brightest points on either side of the disk at 11" separation. The same procedure was repeated for the 1997 image, which gave very similar results. In this figure north is up.} \label{Fig-DiskAngle}}
\end{figure}


Recently, it was pointed out that the \bpic disk midplane is not straight, but slightly bowed \citep[][]{Milli2014}. We test evidence
for the overall bow in our data by independently fitting the two sides of the disk. We used the analysis quality reduced images from the 2012 observations and fit two lines starting from the position of the star to the brightest point in the disk at the edge of the field of view {\change ($\sim$11" on each side, same points as used above for the disk position angle determination, see black lines in Fig.~\ref{Fig-DiskAngle}).  These fits yield about 0.23$^\circ$ {\em difference}  between the position angle of the two sides. As illustrated in the inset of Fig.~\ref{Fig-DiskAngle}, the star's position lies NE from the vector connecting the SW and NE disk midplane (white line), which is consistent with a curvature to the NW direction, i.e. the two disk wings slightly bowed to the SE direction.} In other words -- with respect to a straight line fitted on the disk -- the inner disk appears slightly NW of the line and the two disk wings (in NE and SW directions) both are slightly bent to the other side (SE).
The fact that the disk is slightly bowed may be due to the interactions with ISM \citep[e.g.][]{Gaspar2008,Buenzli2010} or, more likely, due to the effect of forward-scattering grains in a not completely edge-on disk \citep[][]{Rodigas2014}; given the direction of the curvature and the inner disk asymmetries, it is not likely that the overall bow is a result of radiation pressure-driven grains emerging from the asymmetric inner disk.

\subsection{Surface Brightness Along the Disk Midplane}
\label{DiskSurfaceBrightnessProfiles}

To measure the disk surface brightness profile we first rotate the "north up", PSF-subtracted image of the disk (by 60.9$^\circ$ CCW) to place the main axis of the disk on the image horizontal and then extracted a 4 pixel-wide rectangular stripe centered on the disk mid-plane. The image was converted to physical flux density units by multiplying by the STIS inverse sensitivity (4.0169$\times10^{-19} $ergs/s/cm$^2$/\AA \, per count/s for the 1997 image and $4.1446\times10^{-19}~$ergs/s/cm$^2$/\AA \, per count/s for the 2012 image), a value that is provided by STScI's and is based on the absolute flux calibration of the STIS instrument.  Next, for each radius (at each pixel) we calculated the median of the vertical stripe, excluding pixels not sampled in the image. 

We applied this process to the northeastern and southwestern wings of the disk and for both epochs (1997 and 2012), 
resulting in four surface brightness profiles. Figure~\ref{Fig-Profiles} shows the resulting surface brightness profiles for the 1997 and 2012 data.The 2012 STIS observations have closer inner working angle and probe the disk surface brightness all the way down to 0\farcs35; however, we consider the data reliable (i.e. sampled by
multiple pixels at a given radius) only for radii greater than 0\farcs4.  

{\change In Fig.~\ref{Fig-Profiles} we also include representative $\pm1\sigma$ uncertainties for two of the profiles as blue and red shaded envelopes. We estimated the uncertainties at each radial bin as the standard deviation within the 4-pixel wide stripe, which reflects the combination of photon and readout noise (minor contributions), PSF subtraction residuals, and actual vertical differences in the surface brightness in the disk (major contribution).  The fact that the actual scatter in our profiles is smaller than the envelope suggests that our uncertainty estimate is conservative. We also considered another component, HST's photometric calibration uncertainty, but concluded that it is not significant for our analysis for the following reasons. The calibration uncertainty is less than 1\% (see \S,~\ref{S:TemporalChanges}) and would therefore not visibly change our current noise estimate. In addition, it would effect all measurements the same way (small relative increase in surface flux density), i.e. and would therefore not change our results on the surface brightness slopes (see below). 

In Fig.~\ref{Fig-Profiles} we also plot in light gray the ALMA 1.3~mm dust continuum profile (integrated in the direction perpendicular to the disk midplane) from the measurements presented in \citet[][]{Dent2014}. The mm-sized grain population traced by the ALMA observations is thought to follow the distribution of the planetesimals in the disk, but it is somewhat complicated by the projection effect in the nearly edge-on viewing geometry. We point out that the regions R2--R6 -- covering annuli between break points in the surface brightness slopes -- appear to all coincide with inflection points in the dust continuum. This agreement suggests that the fine grains seen in scattered light provide also approximately trace the distribution of planetesimals.}

\subsection{Large-scale Radial Asymmetry} 

We confirm a previously reported brightness difference between the NE and SW wings of the disk. The SW side of the disk is brighter in the inner {\change 8\farcs0} and fainter at larger distances on the NE side (see Fig. \ref{Fig-Profiles}). We find that this brightness difference extends to the previously unexplored inner disk down to at least 0\farcs5 or about 10~au projected radius. The brightness difference is greater at the smallest separations ($\sim$50\%), initially decreases toward larger radii, and the two sides reach identical brightness at 2\farcs5. Between 3\farcs0 and {\change 8\farcs0} the SW side is again brighter, but its brightness drops faster beyond 7\farcs0 than that of the  NE side and beyond {\change 8\farcs0} the NE side is seen to be brighter. 

We note that the more precise characterization of the surface brightness profile is limited by the fact that the disk has considerable vertical structure (\S~\ref{VerticalStructure}) and thus projecting it to a one-dimensional distribution is an ill-defined problem: different vertical aperture widths will return slightly different results depending on what disk regions are included.

\subsection{Breaks in the Radial Surface Brightness Profile}
\label{RSB}

\begin{figure}
\epsscale{1.10}
\plotone{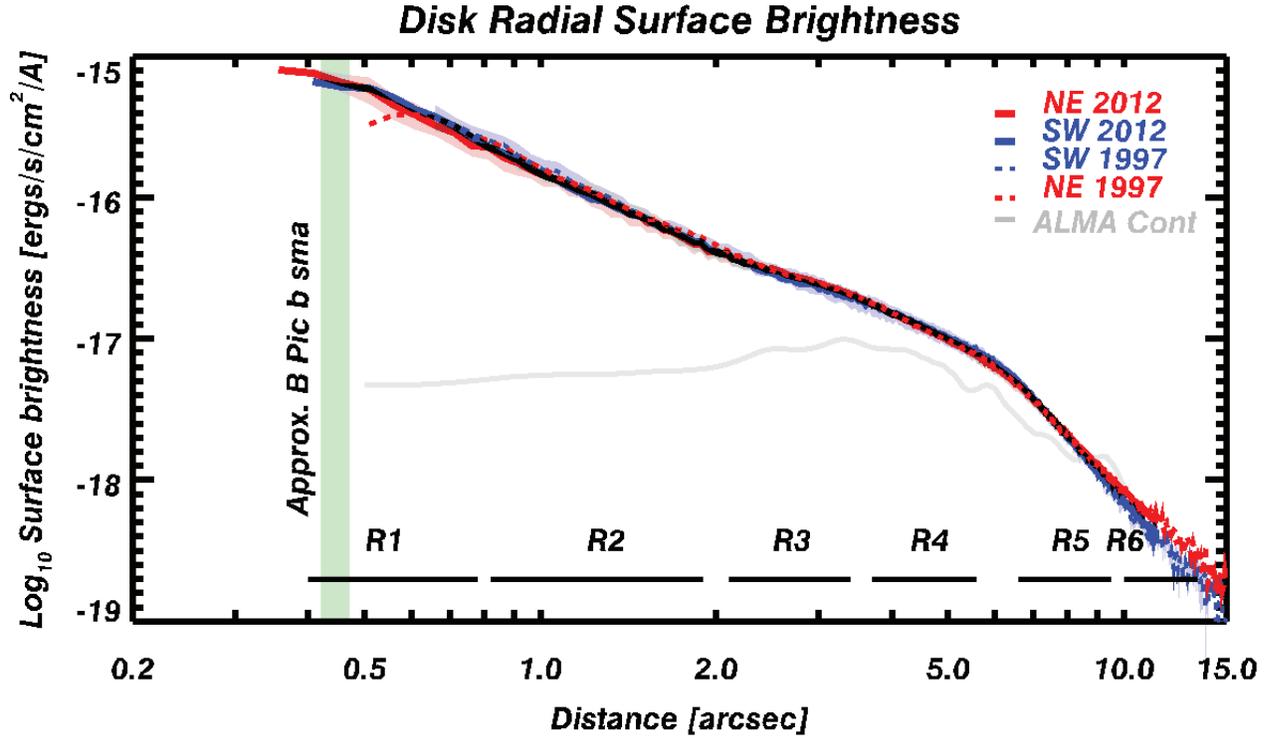}
\caption{Comparison of the disk surface brightness profiles from two epochs and two sides of the disk. The disk shows an NE-SW asymmetry: the NE (red) side is slightly fainter in the inner {\change 8\farcs0}, but brighter at larger separations than the SW (blue) side. Multiple breaks are present in the disk surface brightness: at 2\farcs0 in the SW wing only and at 6\farcs0 in both wings. The profiles observed in the two epochs show near-perfect matches, except for a possible fainter emission in the newer {\change (2012, red dashed line) NE} measurements at radii between 1\farcs0 and 1\farcs7".  {change The light gray curve shows the ALMA 1.3mm continuum profile (Dent et al. 2014) integrated perpendicular to the disk and presented here in a linear, normalized scale. Note, that several of the disk region boundaries identified in the scattered light image the correspond to inflection points in the planetesimal distribution traced by the ALMA continuum data.} The shaded region marks the probable semi-major axis range of $\beta$ Pic~b. \label{Fig-Profiles}}
\end{figure}

The radial surface brightness profiles show distinct disk regions with
different slopes. To characterize the surface brightness profiles we fitted linear slopes to the logarithms of the 
projected separation and the surface brightness profiles (see Fig.~\ref{Fig-Profiles}). 
We find that the slopes are very well constrained and very similar in the two epochs. 
We note that the uncertainty of the slopes is not dominated by noise in the images but by the precise definition of the aperture over which the slopes are calculated in the non-linear and extended disk. We estimate the
typical uncertainty of the power-law coefficients to be $\pm$0.04.

Our analysis of the disk's radial surface brightness reveals at least five distinct disk regions 
(see Table~\ref{TableRadialProfiles}) ranging from 0\farcs4 to 13\farcs4, corresponding to 7.7 au to 260~au. 
We labeled these R2--R6 and also included region R1, which is only probed in our 2012 image.
We characterized the surface brightness slopes by fitting the brightness distributions in log-log space with
linear functions, corresponding to $S\propto r^\alpha$ power-law radial distributions. 
The surface brightness slopes determined range from a mild slope in the inner disk ($\alpha \simeq-1.7$ to $-1.8$ for
radii smaller than 39 au, {\change i.e. Regions R1 and R2}), surrounded by an even milder slope ($\alpha \simeq-1.4$ for radii between 39 au and {\change 68 au, Region R3}).
Beyond 70 au we find three disk regions with much steeper slopes: the first has $\alpha \simeq-1.9$ (between 72 au and 110 au, Region R4), the following large region shows the steepest slope observed ($\alpha \simeq-4.4$ for radii between 130 au and 190 au, Region R5). The outermost disk region again shows a steep, but somewhat milder slope ($\alpha \simeq-3.9 $ to $-4.5$ for radii between 190 au and 260 au, Region R6).

The disk surface brightness profile has been studied in detail by \citet[][]{Golimowski2006} on the basis of
their HST/ACS F606W coronagraphic observations. The boundaries of our regions R3--R6 were matched to 
those of the four regions identified by \citet[][]{Golimowski2006} (regions 1--4 in their paper) allowing 
direct comparison  (see Column 4 in Table~\ref{TableRadialProfiles}). {\change We note here that we adopted the slopes \citet[][]{Golimowski2006} derived from the non PSF-deconvolved ACS images; these show greater similarity to our STIS images than the PSF-deconvolved ACS images of \citet[][]{Golimowski2006} .}
We find that the overall radial disk structure  {\change observed in our STIS images matches closely} that described by those authors. 
In spite of the good overall agreement the surface brightness slopes fitted to the two different datasets are slightly different,
which we attribute to the different spectral coverage of the unfiltered STIS observations ($\sim$200-1,050 \AA) and the filtered 
F606W ACS images. 

Our 2012 images have smaller inner working angles than the previous ACS and STIS observations, allowing us to accurately measure the radial surface brightness of the inner disk. In the 2012 data set we can probe the disk structure down to 0\farcs4. We find that the slope observed between 0\farcs8  and 2\farcs0 in the SW disk continues inward in the NE disk without significant change down to radii 0\farcs4.

\begin{table}
\begin{center}
\caption{Power-law fits to radial surface brightness profiles and comparison to the values measured by Golimowski et al. (2006)  (see their Table 3). \label{TableRadialProfiles}}
\begin{tabular}{cccrrr}
\tableline\tableline
Region &Range & Range & \multicolumn{3}{c}{$\alpha$}    \\
&[arcsec] & [au] &1997 STIS & 2012 STIS & F606W ACS\tablenotemark{a} \\
\tableline
Northeastern Disk\\
\tableline 
R1  & 0.4-0.8 	& 7.8--16  &   $-1.17\pm0.43$  &  $-1.80\pm0.20$  		&  --- \\
R2  & 0.8-2.0 	& 16--39	&  $-1.80\pm0.05 $             & $-1.87\pm0.10 $   	&  --- \\
R3  & 2.0-3.5 	& 39 -- 68	&  $-1.36\pm0.06  $           & $-1.26\pm0.07   $ 	&  $-$1.34 $\pm$0.02      \\
R4  & 3.7-5.6 	&  72 -- 110   	&  $-2.05\pm0.06  $            & $-1.99\pm0.07 $   	& $-$1.91$\pm$0.02        \\
R5 & 6.6-10.0 	& 130--190	&  $-4.23\pm0.07 $            & $-4.19\pm0.08		$& $-$4.19$\pm$0.02        \\
R6 & 10.0-13.4 	& 190 -- 260 &  $-4.25\pm0.57 $             & $-3.92\pm0.49$\tablenotemark{b}        & $-$3.63$\pm$0.02        \\
\tableline
Southwestern Disk\\
\tableline
R1 & 0.4-0.8 		& 7.8--16   & $-1.91\pm1.63 $& $-1.58\pm0.33$ & --- \\
R2&0.8-2.0 		& 16--39&  $-2.00\pm0.11 $   &  $-1.86\pm0.11 $  &   ---      \\
R3&2.0-3.5 		&  39--68&$-1.45\pm0.16 $   &  $-1.37\pm0.16 $  & $-$1.06$\pm$0.02        \\
R4&3.7-5.6 		& 72--110  & $-1.89\pm0.18 $&  $-1.85\pm0.18$  & $-$2.03$\pm$0.02        \\
R5&6.6-10.0 		& 130-190 & $-4.83\pm0.09$  &  $-4.70\pm0.09$    & $-$4.76$\pm$0.02        \\
R6& 10.0-13.4 	& 190--260 & $-5.00\pm0.52$     &  $-4.36\pm0.63$\tablenotemark{b}  & $-$4.00$\pm$0.02        \\
\tableline
\end{tabular}
\tablenotetext{a}{The HST/ACS uncertainties may not include the systematics fully.}
\tablenotetext{b}{The outer radius of Region R6 in the STIS 2012 images is 11\farcs0. }
\end{center}
\end{table}

\section{Vertical Disk Structure and Asymmetries}
\label{VerticalStructure}

The \bpic disk shows a complex vertical-radial structure noted since after the disk was first imaged
\citep[e.g.][]{SmithTerrile1984,KalasJewitt1995,Heap2000,Golimowski2006}.
 Our high-quality and small working angle images allow detailed study of the inner disk structure. In the following we use different analysis techniques to probe the asymmetries
in the disk. First, we show and discuss a radially normalized disk image and model the vertical structure
at each radius as a sum of two gaussian functions. Second, we will explore the rotationally asymmetric disk
component. Third, we attempt to separate the main disk contribution from the warped component. 

\subsection{Radially Normalized Vertical Structure}
\label{Section:VerticalStructure}
\begin{figure}
\epsscale{0.8}
\plotone{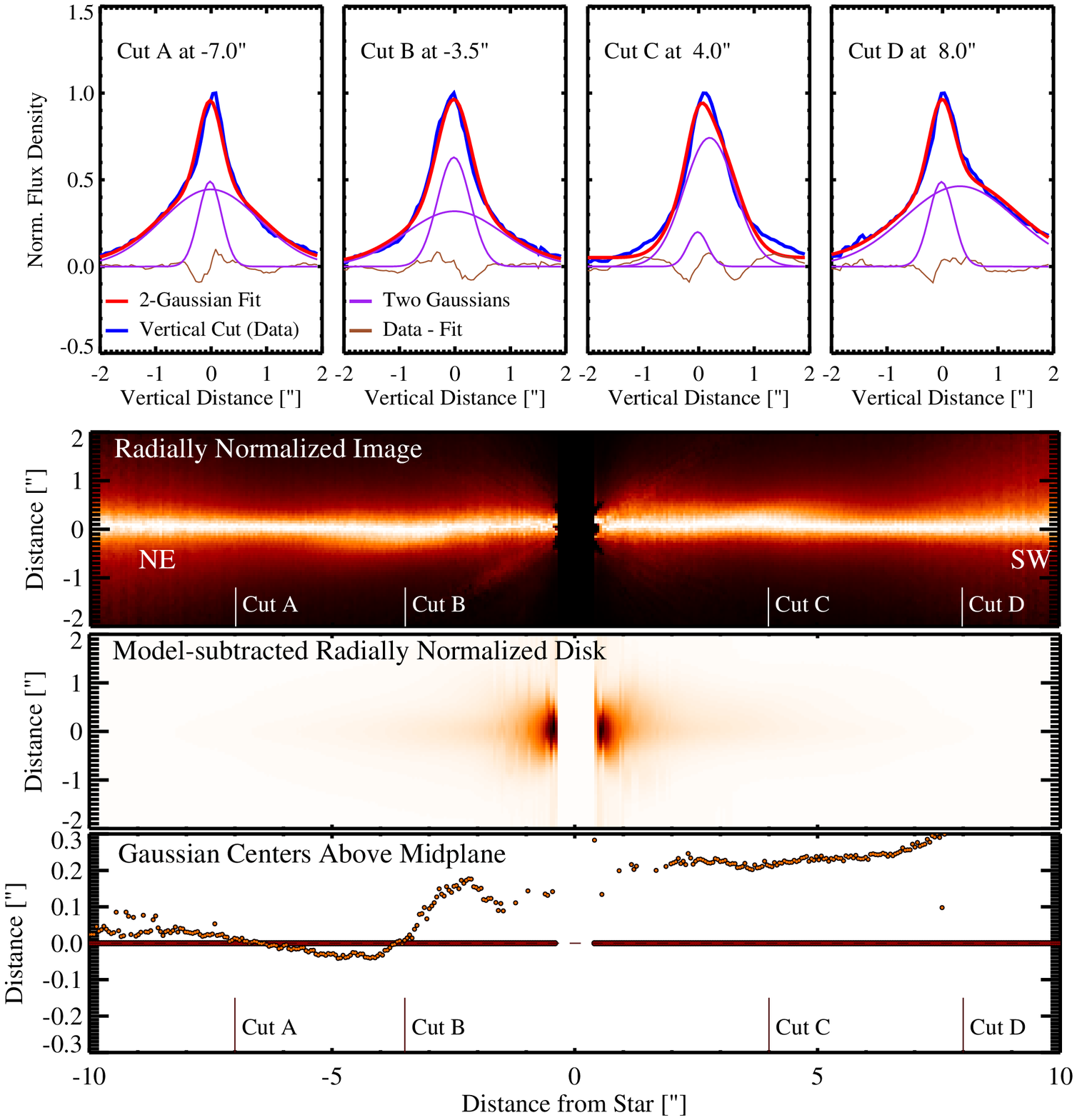}
\caption{{\em Top panel:} Disk vertical profiles (blue) at Cuts A, B, C, and D fit by a sum of two Gaussians (sum in red, components in purple). The center of one of the Gaussians was fixed to the disk midplane, but all other parameters were unconstrained. {\em Second panel:} A radially normalized disk image shows a complex vertical structure. The locations of cuts A, B, C, and D are marked. {\em Bottom panel:} The vertical offset of the first- and second Gaussian fits to the vertical disk profiles as a function of radius. The plot reveals large NE-SW asymmetry, mainly introduced by the presence of the warp (within 4"). The first gaussian fits are the darker symbols.  \label{RNormalized}}
\end{figure}

\begin{figure}
\epsscale{0.8}
\plotone{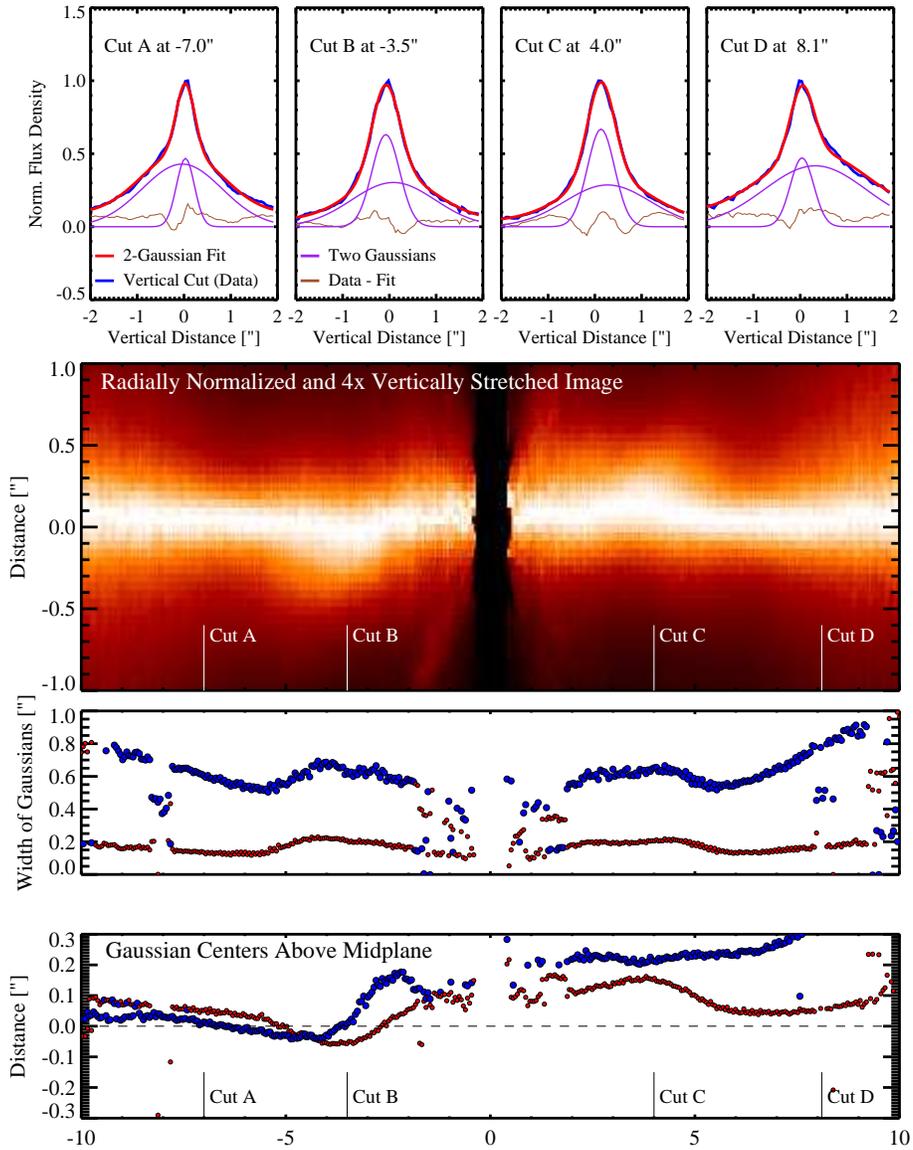}
\caption{{\em Top panel:} Same as in Fig.~\ref{RNormalized}, but all parameters of the two Gaussian were unconstrained and the radially normalized disk is shown in a 4$\times$vertical stretch.  \label{VertExpansion}}
\end{figure}

We now inspect the disk vertical structure by modeling it with analytical functions at each radius. To better 
visualize the vertical structure we normalize the disk brightness at each radius (1 pixel vertical slice) by the 
peak value at each corresponding radius (see panel  with {\em Radially Normalized Image} in Fig.~\ref{RNormalized} and a 4$\times$ vertically stretched version in Fig.~\ref{VertExpansion}). This image reveals a prominent asymmetry in the disk: the disk is  warped in its NE (left in Fig.~\ref{RNormalized}) wing between $\sim2\farcs$ and $\sim5"$; a similar, but not identical warp is apparent in the SW wing between $\sim3\farcs0$ and $5\farcs5$.
The warp, in effect, introduces a counter-clockwise deformation in the disk, i.e. adding excess scattered light
south of the NE wing and north of the SW wing. 

To quantify the vertical structure we fitted each vertical slice with the combination of two Gaussian functions. In Fig.~\ref{RNormalized} we fixed the center of one of the Gaussians to the disk mid plane, while in Fig.~\ref{VertExpansion} we treated the centers of both Gaussians as free parameters.  The
fits were optimized by the IDL function {\tt MPFIT} \citep[][]{Markwardt2009} which applies Levenberg-Marquardt least-squares optimization.
We found that the sum of two Gaussians provided a very good fit for all radii, with the exception of the innermost
1". The top panels in Figs.~\ref{RNormalized} and \ref{VertExpansion} show the vertical disk brightness
profiles (blue), the best-fit analytical functions (red), and the residuals (thin black lines) for the four vertical slices
shown in the radially normalized image (labeled as Cuts A--D). 

These best-fit analytical functions allow us to generate a simple "disk model", which we subtract from the
actual radially normalized image to verify how well this simple approach can reproduce the disk structure.
In the second panel from the bottom of Fig.~\ref{RNormalized} we show the model-subtracted radially
normalized disk. This image demonstrates that our simple analytical fit captures well most of the disk structure down to a stellocentric separation of 1".
We note that the fit shown in Fig.~\ref{VertExpansion},  {\change in which all parameters of both Gaussians are unconstrained}, provides a better fit to the disk than the more constrained fit shown in Fig.~\ref{RNormalized}. 

The bottom panels in Figs.~\ref{RNormalized} and \ref{VertExpansion} show the offsets of the centers of the two Gaussian functions relative to the disk midplane. The four cuts (A--D) are also marked on this plot. The third panel in Fig.~\ref{VertExpansion} shows the width of the two Gaussians. 

{\change As a cautionary note, we stress here that the two-Gaussian fit is not physically motivated and does not  capture the true complexity of the disk. The two Gaussians do not provide a one-to-one match to the structures an observer would identify as the main disk and the warped disk (or primary and secondary disks). Nevertheless, the combination of the two Gaussians provides an overall good fit to the {\em entire vertical} disk profile and allows us to trace the general radial dependence of the vertical structure, without decomposing the disk to physical components. }

\begin{figure}
\epsscale{1.1}
\plotone{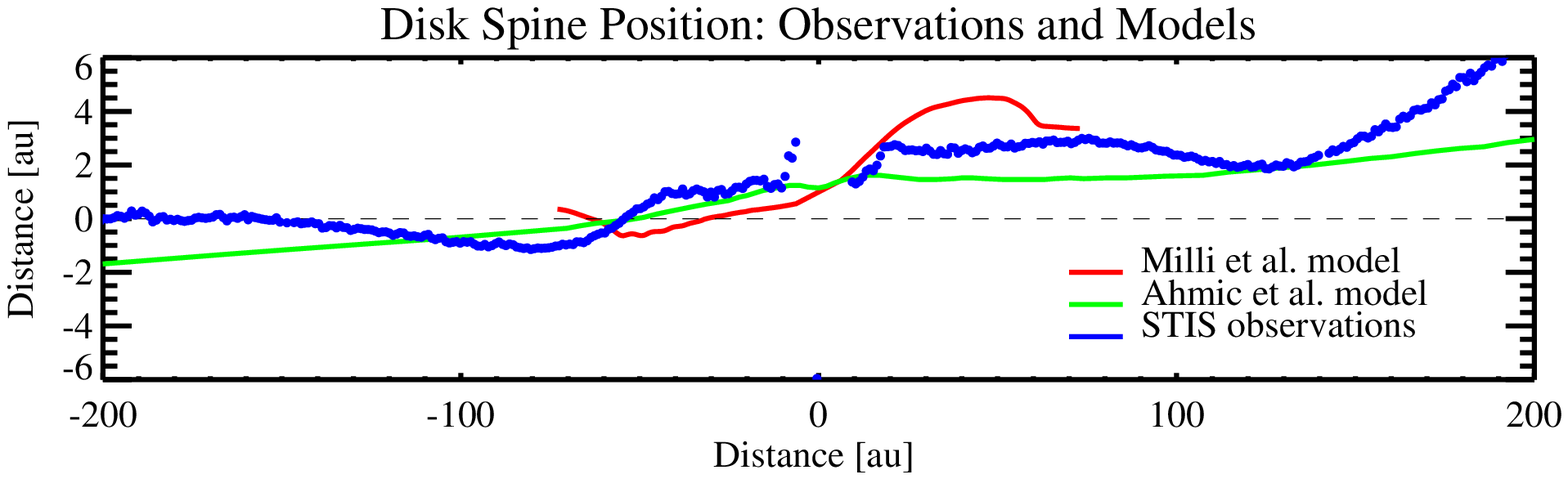}
\caption{\change Comparison of the observed and modeled vertical position of the disk spine (brightest point). Shown are composite (two-disk) models by \citet[][]{Milli2014} and \citet[][]{Ahmic2009}. The blue symbols show the peak position of a single vertical Gaussian function fitted to the STIS observations presented in this paper, an analogous measurement to the modeled spines shown. \label{Models}}
\end{figure}

The Gaussian centers (shown in bottom panels in Fig.~\ref{RNormalized} and \ref{VertExpansion}) reveal a NE-SW asymmetry and trace the vertically extended disk structure most prominent at $\pm$4". 
This structure appears as a counter-clockwise warp in the disk that is traceable to radii $<$1". The structure seen here has been identified as a "warp" in early STIS images \citep[e.g.][]{Heap2000} {\change and}, based on PSF-deconvolved 2003 HST/ACS images, \citet[][]{Golimowski2006} have {\change described it as an inclined {\em secondary disk}. Recently, \citet[][]{Ahmic2009} and \citet[][]{Milli2014} have modeled the appearance of a two-component dust disk, considering anisotropic scattering and high inclination (but not exactly edge-on). }

At larger separations in the SW wing -- the Gaussian center(s) are again positively offset above the disk midplane and this opening continues beyond the field of view of our images. The NE wing shows a similar, but less pronounced structure. These asymmetries have been recognized in early coronagraphic studies at even larger separations and referred to as "butterfly asymmetry" \citep[][]{KalasJewitt1995}.

\subsection{The Warp: Separating Asymmetric Structures}
\label{SectionWarp}

\begin{figure}
\epsscale{1.10}
\plotone{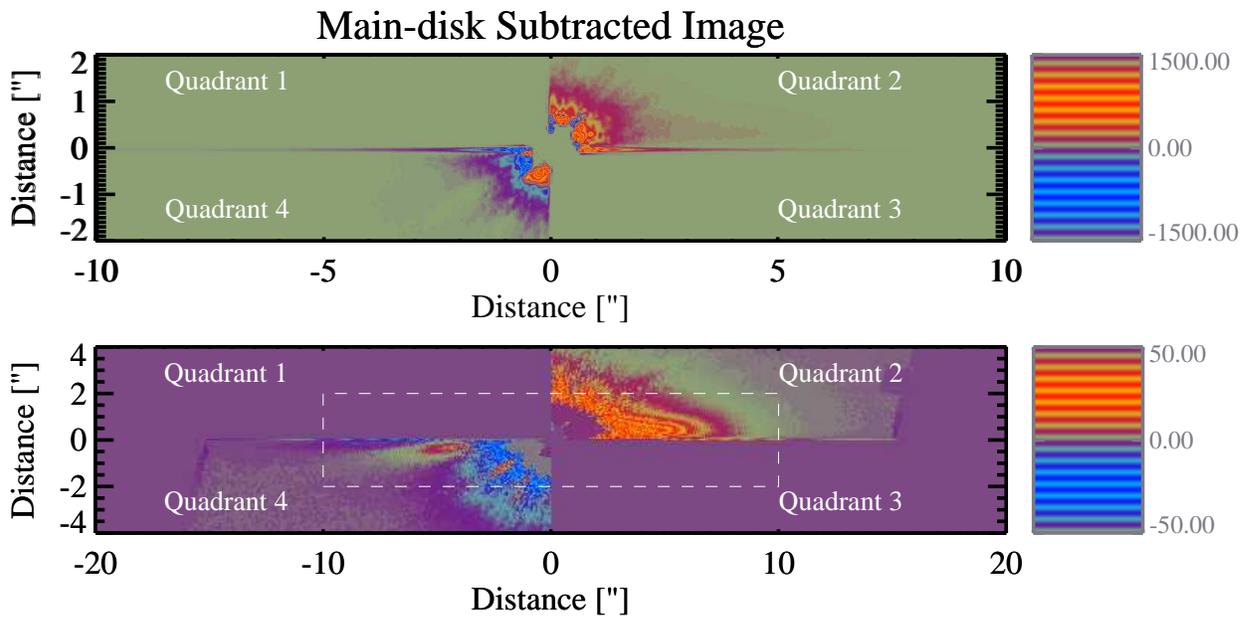}
\caption{Result of the subtraction of the quadrants without warp from the two quadrants with warp. The warp structure shows a clear NE-SW asymmetry. The image is rotated by 60.9$^\circ$ counter-clockwise from the standard north up orientation to align the disk with the x-axis.  \label{Fig-MainDiskSubtracted}}
\end{figure}

We explore the disk warp by subtracting the contribution of the "main disk", based on the assumption that
the main disk has a perfect mirror symmetry to the main disk midplane, following {\change \citep[][]{Lagrange2012}}. Therefore, we identify four quadrants in
the image (numbered clockwise from 1 to 4, see~Fig~\ref{Fig-MainDiskSubtracted}), two of which ({\change 2nd and 4th}) contain large 
flux contributions from the warped disk, while the other two ({\change 1st and 3rd}) are dominated by emission from 
the main disk. 
By mirroring the {\change 1st and 3rd} quadrants to the main disk mid-plane and combining these with the original
{\change 1st and 3rd} quadrant images we can approximate the main disk. We then subtract this combined image from the 2012 data set, thus producing an approximately main-disk-subtracted image (see Fig.~\ref{Fig-MainDiskSubtracted}). 
{\change We note that because the main disk is not perfectly symmetric with respect to the disk midplane this method is only providing an imperfect separation of the warped structure. }

This subtraction highlights the disk warp and its NE-SW asymmetry. The location and the asymmetry are consistent with those seen in Fig.~\ref{VertExpansion}: in the NE wing the warp appears around 2$-$4", while in the SW wing it is present from 1" to 5"; not only is the emission more extended in the SW, but it is much more pronounced (i.e. brighter). {\change The mirror-subtracted image shows a morphology that appears to be consistent with a warp caused by an inclined secondary disk (see Section~\ref{Section:VerticalStructure}).}

{\change In Fig.~\ref{Models} we compare the radial dependence of the disk spine -- as measured by the peak position of a single vertical Gaussian function -- to two different recent models by \citet[][]{Ahmic2009} and \citet[][]{Milli2014}. From NE to SW the STIS single-Gaussian fit traces a deviation to "below" the disk mid-plane in the figure, corresponding to the SE direction, between 140 and 55 au, followed by a turn to the other, NW, side of the disk . From NE 55 au toward the star the disk spine is located "above" the disk mid-plane in the figure, corresponding to the NW plane of the disk. The disk spine displays another break around SW 70 au and turns back toward the disk midplane; until 130 au the disk spine is approaching the disk midplane. From 130 au it displays another break and leaves the disk midplane. 

Both models shown as comparison in Fig.~\ref{Models} use a two-component disk with anisotropic scattering phase functions to reproduce the disk spine as seen in projection and in scattered light.
The \citet[][]{Ahmic2009} model has been specifically developed to match the HST/ACS images presented in \citet[][]{Golimowski2006}, which is a data set very similar in wavelength to the STIS data presented here, but does not extend to the inner working angle shown in our images. The model from \citet[][]{Milli2014} has been developed to explain the VLT/NACO L' observations presented in that paper; the VLT paper has a similar inner working angle to the STIS data presented here.

Both papers attempt to reproduce the warp observed in \bpic with a combination of two misaligned dust disks. The \citet[][]{Ahmic2009} study uses two axisymmetric, partly spatially overlapping disks, each with multi-component power-law density distributions. A Markov Chain Monte Carlo routine optimizated the disk parameters to best reproduce the ACS observations. The \citet[][]{Milli2014} study used a grid-based parameter exploration.

We compare our observations to these models as a preliminary exploration to asses whether the two-component disk geometry is consistent with the observed disk spine location. We note that while neither of the two models matches well our new, smaller inner working angle data, they both offer overall similar spine geometries. While our observations do not directly show a secondary disk, the general spine morphology supports the interpretation that the inner disk is inclined with respect to the outer disk, leading to a "warp" as seen in projection. We emphasize that comprehensive scattered light modeling is required for further interpretation of the disk, but such an effort is beyond the scope of the current work.  
}

\section{The Planet and the Inner Disk }
\label{S:Orbit}

In this section we explore possible connections between the disk structure and the super-jupiter
\bpic b embedded in the disk. First,  we compare \bpic~b's orbit to the inner disk structure as seen in our 2012 STIS images.
As of now the orbit of \bpic~b is closely monitored, but not yet fully known. The planet was first imaged in 
Nov. 2003 \citep[][]{Lagrange2009}, but follow-up observations could not confirm its presence until 
2009 \citep[][]{Lagrange2010}. After its confirmation 
the planet was followed by multiple teams and, with over a dozen additional detections, its orbit has
been significantly refined {\change \citep[e.g.][]{Bonnefoy2011,Chauvin2012, Males2014,Macintosh2014,Bonnefoy2014}}. Although the orbit is well constrained after 2009, uncertainties are very large in its position prior to 2009, based on a single detection.  

Here we adopt results from one of the latest orbital analyses of \bpic b, which includes all published and several yet
unpublished VLT/NACO observations {\change \citep{Lagrange2014}, 12 positions in total}. The planet's orbit has been fitted with a Markov Chain Monte Carlo-based (MCMC) optimization of a Keplerian orbit.  
With less than one quarter of the planet's orbit followed accurately,  multiple solutions are possible. Table~\ref{T:Orbitalelements} summarizes three different solutions that are consistent with the existing astrometry.
The three solutions differ in the following ways: In Solution 1 the eccentricity was fixed to zero; Solution 2 represents
the most likely solution from the MCMC program; Solution 3 is a high-eccentricity solution far from the peak of the distribution, i.e. less likely than
the others, but still consistent with the current set of observations. During the preparation of this manuscript an independent orbital modeling based on existing VLT/NACO and new Gemini GPI observations has been presented by \citet[][]{Macintosh2014} and {\change \citet[][]{Bonnefoy2014}}; we note that all three of our solutions are fully consistent with the best fit solution and its uncertainties given by these papers.

\begin{figure}
\epsscale{0.7}
\plotone{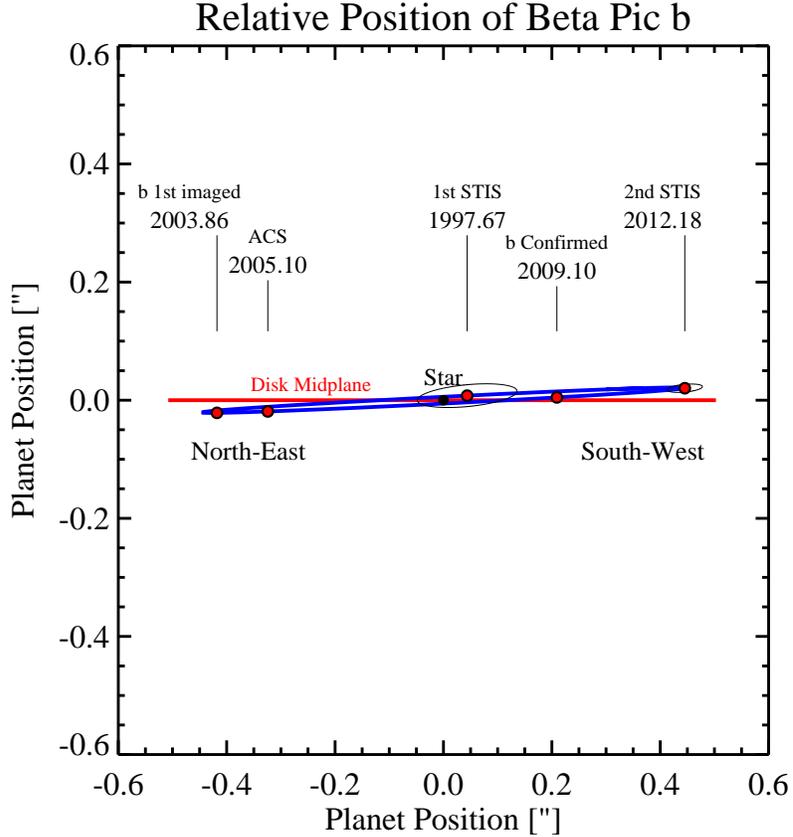}
\caption{The projected orbit of $\beta$~Pic~b (blue) and the disk midplane (red). {\change Selected epochal positions of the planet at are marked
with red symbols}. The orbit shown here corresponds to Solution \#1 in Table~\ref{T:Orbitalelements}. {\change The ellipse corresponds to the
1$\sigma$ positional uncertainty of the planet at the time of the 1997 STIS observations.} \label{Fig-PlanetOrbit}}
\end{figure}

In Figure~\ref{Fig-PlanetOrbit} we plot the best-fit orbit of the planet (Solution \#1) as projected on the plane of the sky. The plot is centered on the star's position and we marked the planet's position at a few significant epochs, including March 6, 2012,  when our HST/STIS observations were obtained. For the 1997 position we also plotted an
uncertainty ellipse with its 1~$\sigma$ semi-major axis set to match the difference in the positions predicted by Solutions \#1 and \#2.

\begin{table}
\begin{center}
\caption{Orbital solutions for \bpic b. \label{T:Orbitalelements}}
\begin{tabular}{ccccl}
\tableline\tableline
Parameter & Solution 1 & Solution 2 & Solution 3  \\
\tableline
a  [au]        & 8.6686      & 9.341403  & 11.18887     \\
P  [yr]          &19.29343   &  21.58237  & 28.29178  \\
e                 &0.0 &  5.160$\times^{-2}$ & 0.163\\
inc    [$^\circ$] &89.269 & 88.603 & 88.803  \\
$\Omega$ [$^\circ$] & $-$148.2210 &$-$148.2013 & $-$147.7354  \\
$\omega$ [$^\circ$]    &  169.5635 &$-$127.3659 &3.996902\\
t$_p$           &2002.229   & 2005.643 &  2013.290\\
$\chi^2$       & 6.26   &  8.67 & 5.367  \\
\tableline
\end{tabular}
\end{center}
\end{table}

\begin{figure}
\epsscale{1.0}
\plotone{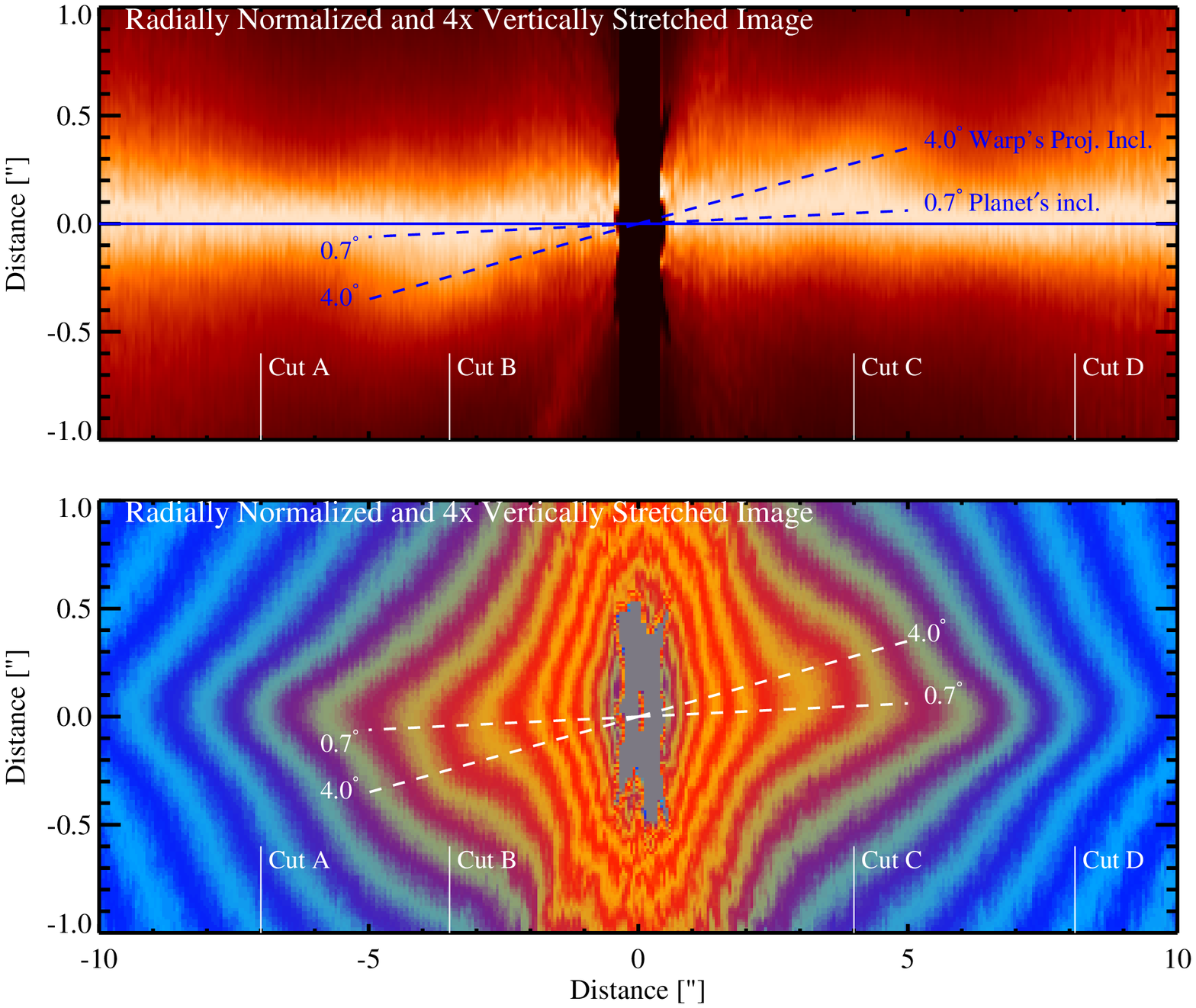}
\caption{ {\change The warp and the planet's orbit in $\beta$~Pictoris. The planet's inclination with respect to the disk midplane is ($0.7^\circ \pm 0.7$, see \S~\ref{S:Orbit}), while the warp (vertically extended disk emission) is seen up to at least 4$^\circ$ from the midplane. } \label{Fig-WarpAngle}}
\end{figure}

\begin{figure}
\epsscale{1.00}
\plotone{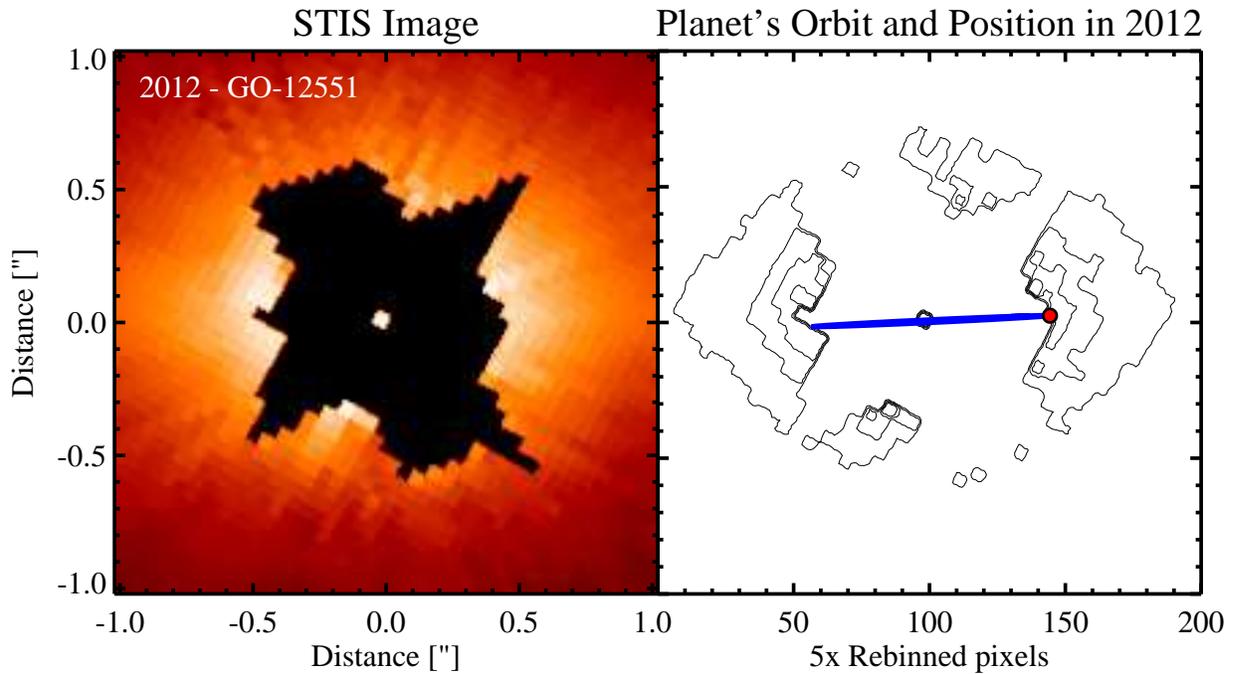}
\caption{{\em Left panel:} The view of the inner disk (2"$\times$2"). {\em Right panel:} The orbit
of the planet (blue) overlaid on a contour plot of the inner disk with the red symbol marking
the position of the planet during the observations. Our images do not show evidence for excess
emission from the planet, but allow probing the disk structure at the orbit of the planet.
{\change The image is rotated by 60.9$^\circ$ counter-clockwise from the standard north up orientation to align the disk with the x-axis.}
\label{Fig-InnerDiskPlanet}}
\end{figure}

In the left panel of Figure~\ref{Fig-InnerDiskPlanet} we show the innermost section of the disk (39$\times$39 au)
centered on the location of $\beta$~Pic. 
Black pixels mark locations covered by the coronagraphic wedge in
every spacecraft orientation and thus without valid data. The image shown here is from our 2012 dataset; the earlier
1997 image has significantly larger inner working angle and it is less informative about the inner
disk structure. 

In the right panel of Figure~\ref{Fig-InnerDiskPlanet} we compare the orbit of the planet with a
contour plot of the inner disk surface brightness. The planet's predicted position for the epoch
of the observations coincides with the innermost valid pixel of our STIS data in the direction of the
planet. However, no evidence for excess emission is seen in the pixel where the planet's point
spread function's core should fall. This suggests that either the planet's optical brightness 
is much weaker than that of the disk or the current best estimate for the planet's orbit is slightly ($\sim0\farcs025$) inaccurate and the planet may be just within our inner working angle. 

The fact that the planet is not visible in our images should not come as a surprise:
Exoplanet atmospheric evolution models from \citet[][]{Baraffe2003} predict a visual  apparent brightness for \bpic b between V=21.2 and V=22.9 (thermal emission only, assuming a mass of 10~M$_{Jup}$, adopting the distance of \bpic, for an age between 8 and 21~Myr \citealt[e.g.][]{Ortega2002,Song2003,Ortega2004,BinksJeffries2014}) and, based on simple geometric considerations, we estimate the reflected light of the planet to be fainter than V=26 (assuming a radius of 1.3~R$_{Jup}$, isotropic scattering, and a Bond albedo of 0.4, 50\% illumination). The predicted brightness from thermal emission and reflected light is much lower than that of the disk at the same radius: from our 2012 image we measure 
7.02$\times10^{-16}$ ergs/s/cm$^2$/\AA ~ or V=16.74 for the pixel to which the planet is projected. This value must be a combination of scattered light from the disk, and thermal emission and reflected light from \bpic b. The fact that this pixel has very similar brightness to the surrounding pixels, which only include the light scattered from the disk, argues for the flux density observed at the location of the planet also being dominated by scattered light emission from the disk. The values discussed here show that the disk surface brightness overwhelms the emission from \bpic by about 5 magnitudes. Note, that the above comparison is only an approximation that does not correctly account for the actual bandwidth of our unfiltered STIS observations; nevertheless, because the choice of V-band for the comparison provides the {\em most favorable } planet--to--disk contrast, a more realistic estimate should find that the contribution from the planet is even lower.

{\change 
In Fig.~\ref{Fig-WarpAngle} we marked the planet's inclination ($\approx0.7^\circ \pm 0.7^\circ$ with respect to the disk midplane, e.g. \citealt[][]{Lagrange2014,Macintosh2014}) and the approximate angle under which the warp is observed in our STIS images ($\approx4.0^\circ$). We note that the warp angle derived from our images agrees very well with the angle of the secondary disk structure described by \citet[][]{Golimowski2006}. The warped disk -- as seen in projection -- subtends an angle larger than the best-fit orbital inclination of \bpic b, suggesting that planetesimals may be perturbed to higher inclinations than that of the perturbing giant planet. This finding is consistent with the predictions of dynamical simulations of a planetesimal system influenced by secular perturbations of a planet on inclined orbits: for example, a study by \citet[][]{Dawson2011} finds that the inclinations of the planetesimals will oscillate between $0$ and $2~i_p$, where $i_p$ is the inclination of the planet. The fact that the warp is seen at angles $4^\circ$, would taken on face value, then suggests that \bpic\, b's inclination is -- within uncertainties -- underestimated by current measurements and it may be close to $\sim2^\circ$.
Alternatively, \bpic b's inclination may have been damped, i.e. now may be lower than in the past, which would explain a low present-day planet inclination with a more inclined inner disk, but would require the presence of another planet \citep[][]{Dawson2011}.
However, we note that scattered light is an imperfect tracer of the inner disk, and the not perfectly edge-on viewing geometry and anisotropic scattering further complicate its interpretation.}

\begin{figure}
\epsscale{0.65}
\plotone{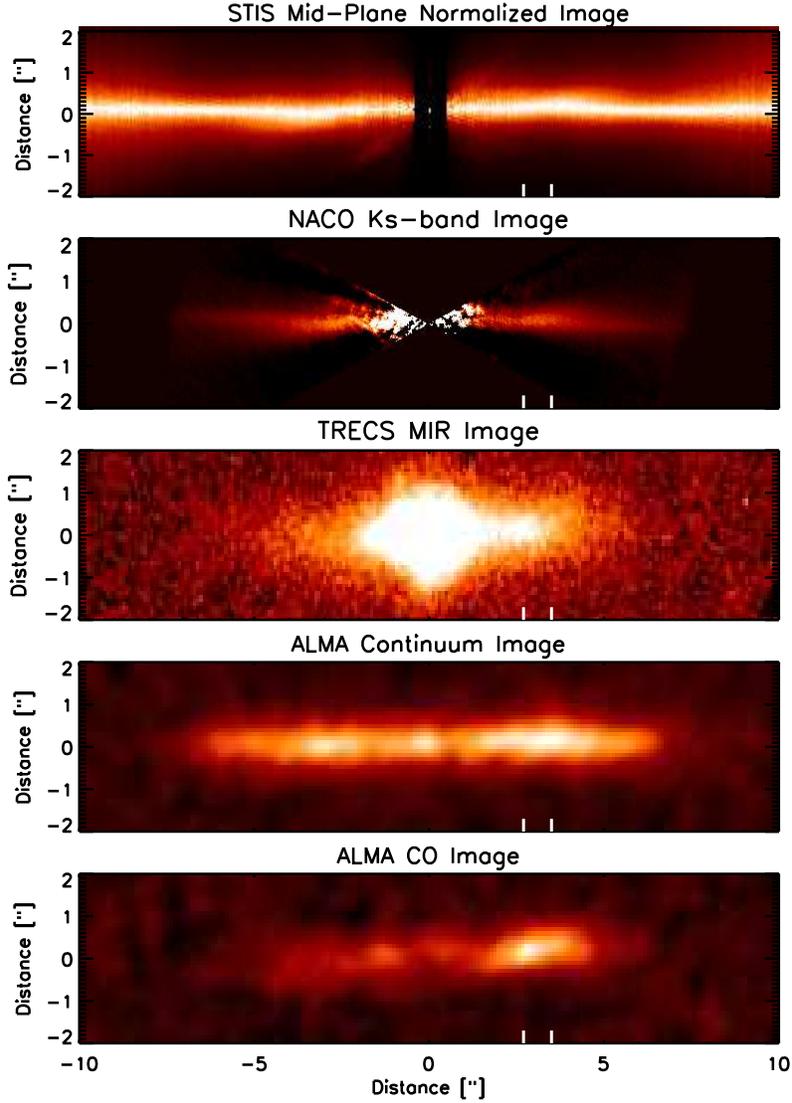}
\caption{ {\change The multi-wavelength view of the disk reveals a major NE-SW asymmetry and a warp. While the NE--SW asymmetry is visible at all wavelengths (from optical through infrared to sub-mm), the warp (emission significantly above the disk midplane) is visible in the optical and near-infrared. The CO peak is also located above the mid plane and may or may not be associated with the warp.  
The left white line marks 2\farcs7, the location of the MIR and CO clumps, and the right white line marks 3\farcs5, the outer edge of the sub-mm dust clump. The blue dot on the top panel also shows the approximate location of the planet during the observations.  Data from this paper, \citet[][]{Lagrange2012}, \citet[][]{Li2012}, and \citet[][]{Dent2014}). {\change The images have been rotated by 60.9$^\circ$ counter-clockwise from the standard north up orientation to align the disk with the x-axis.}}   \label{Fig-Multiwavelength}}
\end{figure}
 
\begin{figure}
\epsscale{1.0}
\plotone{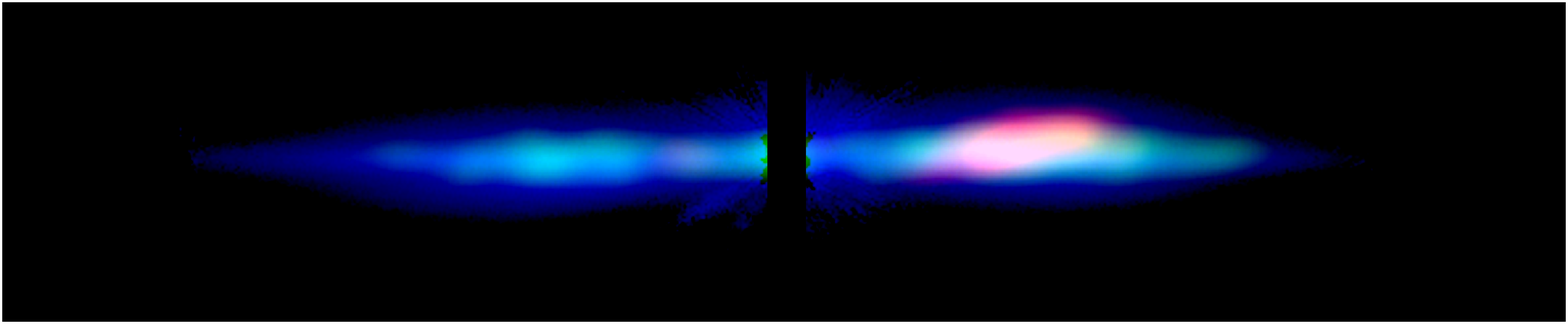}
\caption{ {\change   Color composite image of the $\beta$ Pictoris disk. The scattered light STIS image is shown in blue, the 1.3mm ALMA dust continuum image is shown in green and the CO velocity integrated line emission is shown in red. The ALMA data are from \citet[][]{Dent2014}.  }   \label{Fig-ColorImage}}
\end{figure}

\section{Disk Structure and Its Temporal Evolution}
\label{S:DiskStructure}

\subsection{Multi-wavelength View of \bpic}
The \bpic disk displays multiple radial (Sect.~\ref{RSB}) and vertical (Sect.~\ref{Section:VerticalStructure}) asymmetries, some of which have been proposed consequences of one or two massive planets orbiting within the disk.  In Fig.~\ref{Fig-Multiwavelength} we provide a 
multi-wavelength-view of the disk including optical, near-infrared, mid-infrared, and sub-millimeter images. 
The top panel shows our STIS midplane-normalized image. The second panel from the top shows a VLT/NACO Ks-band image of the disk \citep[][]{Lagrange2012}. For clarity we show the NACO image with two wedge-like sections covering the northern and southern regions directly above the star, areas that have been heavily contaminated by diffraction spike residuals. 

A separate paper in preparation describes the constraints from the VLT/NACO images and we refer to that work for a detailed comparison of the surface brightness slopes; here we will only focus on the disk structure and asymmetries. The Ks-band image traces disk emission in its entire field of view ($\sim14"$ in diameter). Careful analysis by \citet[][]{Lagrange2012} showed that the disk position angle is approximately $29^\circ .3^{+0.22}_{-0.30}$, with the precise value depending on details of the data reduction procedure. These authors also found a slightly different position angle for the SW side of the disk ($\sim209^\circ.10 ^{+0.22} _{-0.38}$); these values are in good agreement (within 1$\sigma$) with the $29.1^\circ\pm0.1^\circ$ value found for our STIS data (Sect.~\ref{DiskAngle}).

The NACO Ks-band image also shows the warp structure seen at optical wavelengths, although less prominently. The disk width, as seen here, reaches its maxima at $\pm$3.5--4\farcs0 separations from the star.
The third panel from the top shows an 11.7~$\mu$m mid-infrared image of the disk emission from \citet[][]{Li2012}. As reported in \citet[][]{Telesco2005} the MIR-clump -- located at 2\farcs7 or 52~au projected separation -- displays a mid-infrared color significantly different from the surrounding disk, indicating that grains in it  differ in temperature, size and/or composition from the particles characteristic of the rest of the disk.

The lower two panels show 870~$\mu$m continuum and $^{12}$CO 3--2 transition images obtained with the ALMA sub-millimeter array \citep[][]{Dent2014}. Both the continuum and the CO image show bright peaks at the SW side of the disk. 
Based on earlier sub-mm images \citet[][]{Wilner2011} argued that the mm peaks at $\pm$3\farcs5 (68 au) mark the locations of a planetesimal belt in which collisions produce large grains. This picture is consistent with the model by \citet[][]{Augereau2001} and also with the new, higher resolution and sensitivity ALMA observations \citep[][]{Dent2014} shown in Fig.~\ref{Fig-Multiwavelength}.

\begin{figure}
\epsscale{0.95}
\plotone{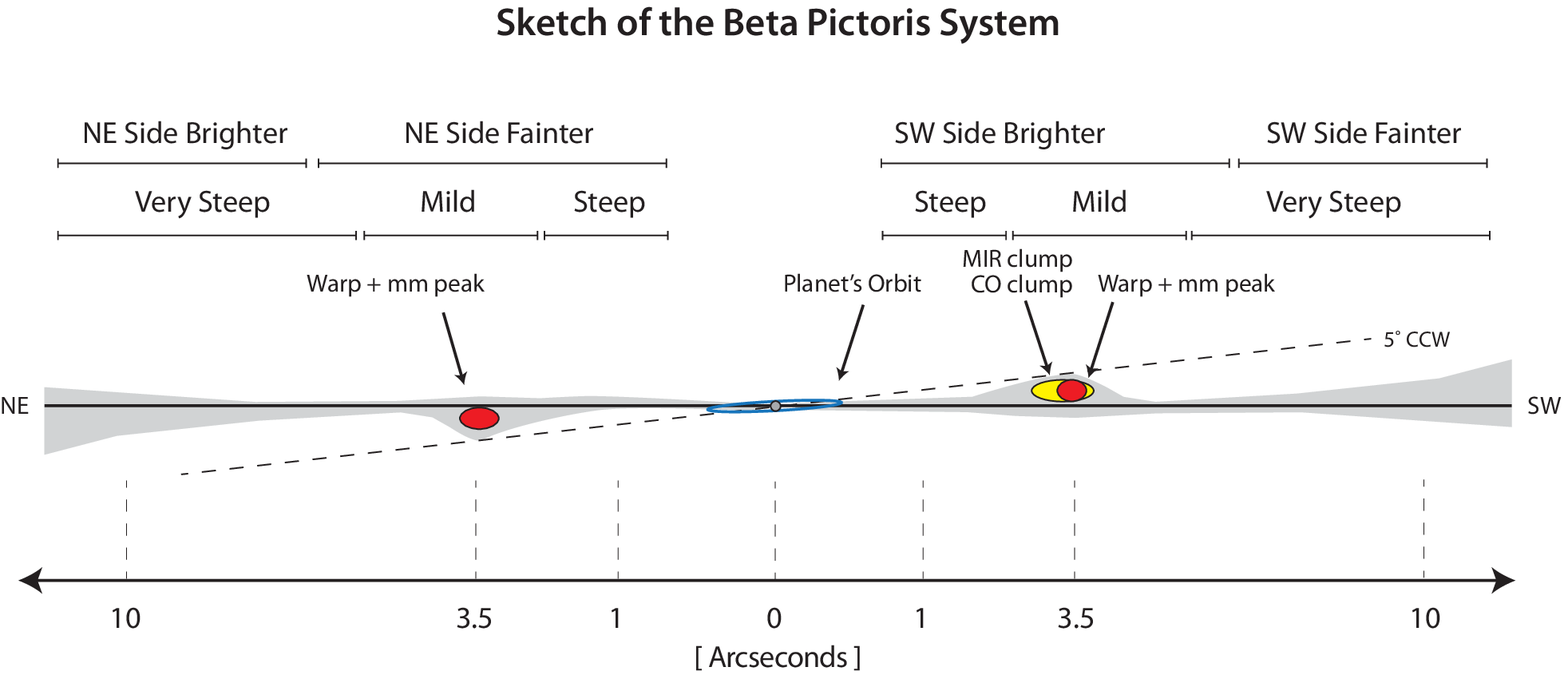}
\caption{{\change Key structures in the \bpic system, as derived from multi-wavelength imaging.} \label{Sketch}}
\end{figure}

{\change Fig.~\ref{Sketch} provides a visual summary of the key structural elements identified in the \bpic disk from multi-wavelength imaging.} Table~\ref{T:Asymmetries} summarizes the asymmetries observed in the \bpic disk at wavelengths ranging from blue optical to sub-millimeter. {\change We also show a color-composite image (Fig.~\ref{Fig-ColorImage}) as an illustration allowing the comparison of the spatial location of the scattered light (blue), the ALMA dust continuum emission (green), and the ALMA velocity-integrated CO emission (red) in the disk.}
We group the observed disk structures in two major categories: apparently axisymmetrical and non-axisymmetrical structures. 

The {\em apparently axisymmetrical}, but not mirror-symmetrical disk structures are primarily those that contribute to the warped inner disk, seen in the optical and near-infrared scattered-light images. There is confident detection of extra emission above (SW) and below (NE) of the optical disk midplane (warp) at {\change 2\farcs5--5\farcs0} at {\em all} wavelengths. 

The {\em apparently non-axisymmetrical} structures are primarily those that reflect the major SW--NE asymmetry: the MIR and CO clumps present in the SW disk, the surface brightness differences between the SW and NE disks, and the different ALMA dust continuum levels. Specifically, there is evidence that the SW side of the disk is brighter than the NE at optical, near-infrared, mid-infrared and sub-millimeter wavelengths. New ALMA observations also show a similar asymmetry in CO gas \citep[][]{Dent2014}. In addition, new Herschel/HIFI observations also argue for a higher abundance of {C~\sc II} gas in the SW wing than in NE wing \citep[][]{Cataldi2013}. The MIR and mm images also argue for a different dust population above the disk, indicating  recently released smaller grains (in the MIR) at a projected separation of 52~au and larger grains that trace the planetesimal belt that released them at 68 au \citep[][]{Telesco2005,Li2012}. Our images also argue for a warp at this location and with the SW warp being brighter than the NE one. Finally, we find that the disk surface brightness slope in unbroken between 0\farcs4 and 2\farcs0, indicating no significant changes in the disk surface density at radii adjacent and exterior to the most likely semi-major axis of the planet.

\begin{table}
\begin{center}
\caption{Summary of the asymmetries observed in the \bpic disk. A visual summary of the key features is given in Fig.~\ref{Sketch}. References: 1 -- \citet[][]{Burrows1995}, 2 -- \citet[][]{Heap2000}, 
3 -- \citet[][]{Golimowski2006}, 4 -- \citet[][]{Wahhaj2003}, 5 -- \citet[][]{Telesco2005}, 6 -- \citet[][]{Li2012}, 7 -- \citet[][]{Lagrange2012}, 8 -- \citet[][]{Weinberger2003}, 9 -- \citet[][]{Wilner2011}, 10 -- \citet[][]{Mouillet1997}, 11 -- \citet[][]{Pantin1997}, 12 -- \citet[][]{Milli2014}, 13 -- \citet[][]{Dent2014}. \label{T:Asymmetries}}
\begin{tabular}{lcccl}
\tableline\tableline
Asymmetry 			 & Radius 	["]    & Proj. Dist. [au]& Fig.  & References \\
\tableline
{\bf Inner Disk}  & & & \\
MIR disk tilted $\sim15^\circ$ CW & $<$1\farcs0     &   $<$19 &        & 4, 8 but see 5 \\
\hline
\multicolumn{3}{l}{\bf Warp -- Axisymmetrical Structures} & \\
Optical: CCW tilt     &  2"-7"&  38--136 &Figs.~\ref{Fig-MainDiskSubtracted} & 1, 2, 3, this work \\
Ks: CCW tilt  & 3\farcs7--4\farcs8 &  70--93 &Fig.~\ref{Fig-Multiwavelength}    & 7\\ 
L$^\prime$: CCW tilt  & 0\farcs4--3\farcs8 & 8 -- 74 & 	&12\\ 
\hline
\multicolumn{4}{l}{\bf SW -- NE Asymmetry -- Non-axisymmetrical structures} \\
SW disk brighter in optical   		& 0\farcs6--8\farcs0  & 12--160 & Fig.~\ref{Fig-Profiles} &  this work\\
SW brighter at 12~$\mu$m              & 1--4"      & 19--38 &      Fig.~\ref{Fig-Multiwavelength} & 11, 4, 9, 5, 6 \\
MIR clump SW side, diff. grains     & 2\farcs7& 52 &Fig.~\ref{Fig-Multiwavelength}   & 4, 5, 6 \\
SW disk brighter in sub-mm 	& 1\farcs5--4\farcs1 & 60  &  Fig.~\ref{Fig-Multiwavelength} & 9, 13 \\
NE disk brighter in optical                 & 8\farcs0-13\farcs0  & 160--630 & Fig.~\ref{Fig-Profiles} &  this work\\
ALMA CO Peaks above midplane & 4\farcs4 & 85 & Fig.~\ref{Fig-Multiwavelength}    & 13\\
\hline
\multicolumn{3}{l}{\bf Bowed Disk: Inclination effect?}  & \\
Slightly curved disk		       & 0--13"  & {\change 0--250} &  Sect.~\ref{DiskAngle}  & 3, 7, 12, this work\\
\tableline
\end{tabular}
\end{center}
\end{table}

\subsection{Temporal Changes and Comparison to Keplerian Motion}
\label{S:TemporalChanges}

We now search for changes between our two images spanning 14.5 years to explore any temporal evolution that may be observable in the disk at optical wavelengths. We followed two different approaches to compare the images: In the first approach we aligned the images based on the stellar position and position angles identified during the reduction process (based on the location of the diffraction spikes and the spacecraft pointing information). In the second approach we searched for the optimal offsets ($\delta$x and $\delta$y) and rotation between the two images by minimizing the residuals in the subtracted image. 
Specifically, we carried out a grid search covering the relative position angle range $+1^\circ$ to $-1^\circ$ (with steps of 0.005$^\circ$) to identify the relative rotation angle that minimizes the sum of the absolute subtraction residuals between the two images at stellocentric separations $>$3.5". For each relative position we cross-correlated the images to identify the optimal offsets. This method has identified 0.0$^\circ$ relative position as the optimal position with offsets of about 1~pixel.
Tests of the second method via the introduction and recovery of artificial image rotations and offsets revealed that while the rotation angle is reliably identified with an accuracy of 0.1$^\circ$ or better, the cross-correlation-based image offset determination's typical uncertainty was $\sim$1.5 pixels (probably set by the broad structures in the images and the noise level). Given that our expected positional uncertainty from the image reduction is more accurate than the cross-correlation based centering, we opted to adopt the first approach outlined above, but note that the second approach also resulted in consistent results. 

In the 14.5 year period separating our Epoch 1 and 2 images the STIS detector response has changed  slightly. The temporal evolution of the detector's response as a function of wavelength has been monitored closely the Space Telescope Science Institute since STIS's activation. The evolution is described by Krist et al. (2013, STIS Instrument Science Report 2013-03). Standard star observations show that in the blue (filter G430L) STIS's sensitivity has declined by 2.8\%, while in the red (filter G750L) its sensitivity has declined by 2.5\%. We used the $F_{Phot,\lambda}$ keywords in the FITS headers of the two datasets to convert our instrumental count rates into units of flux density; these conversion factors reflect the sensitivities calculated for the two different datasets given the unfiltered CCD throughput curves and assuming a flat source spectrum. Given the accuracy of the calibration and the low wavelength-dependence of the sensitivity change we estimate an uncertainty of $\sim$0.3\% in the $F_{Phot,\lambda}$ factor.

\begin{figure}
\epsscale{0.95}
\plotone{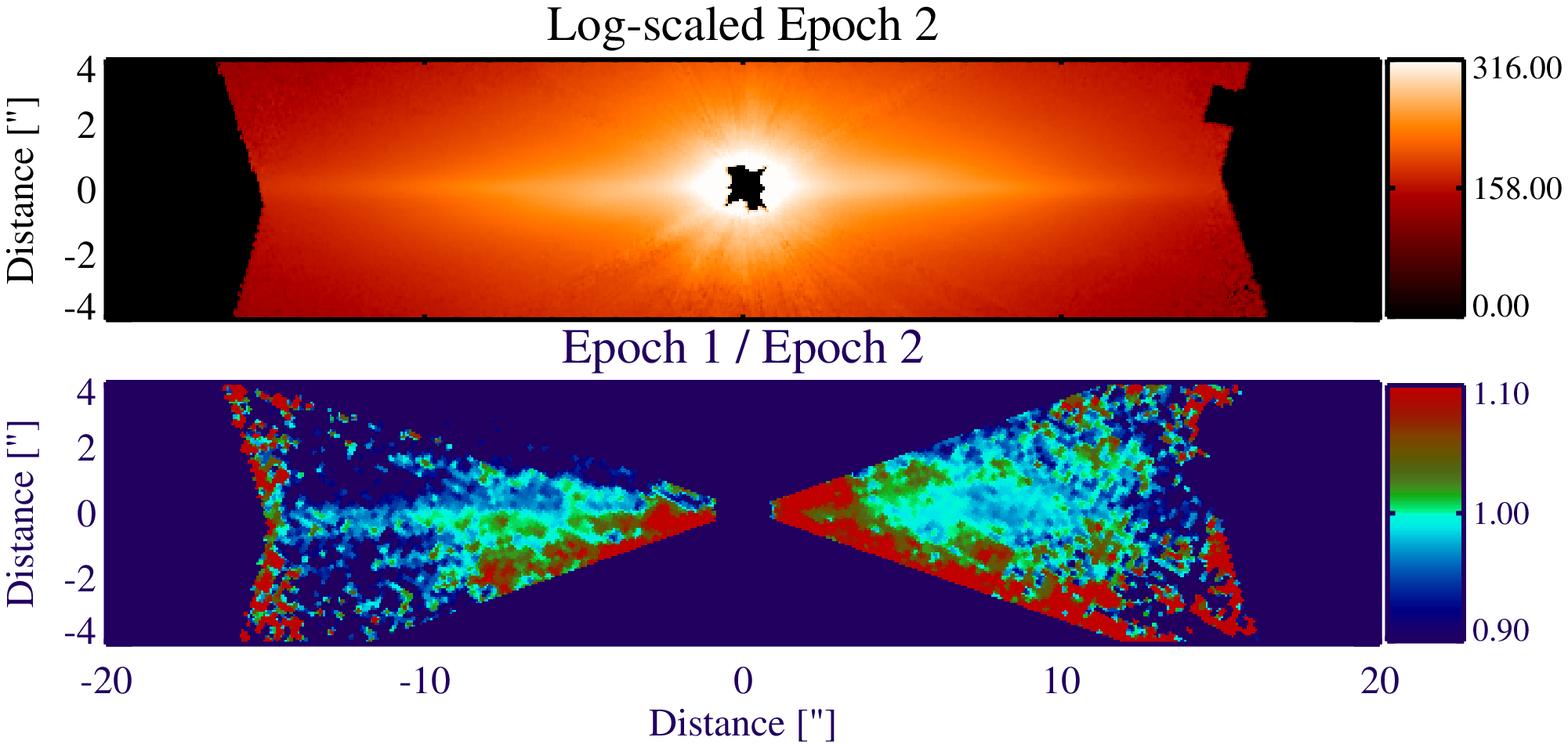}
\caption{ {\change {\em Upper panel:} Logarithmically scaled Epoch 2 disk image. {\em Lower panel:} The ratio of Epoch 1 to Epoch 2 image probes disk surface brightness evolution between 1997 and 2012. The ratio image inside 3\farcs0 is dominated by PSF residuals from the Epoch 1 image. Between 3\farcs0 and 5\farcs5 in the disk midplane no significant changes are observed in the disk structure. The images have been rotated by 60.9$^\circ$ counter-clockwise from the standard north-up orientation to align the disk with the x-axis. }  \label{Fig-Difference}}
\end{figure}

Figure~\ref{Fig-Difference} shows the Epoch 2 (2012) STIS image in a logarithmic scale (upper panel) and the ratio of the Epoch 1 and Epoch 2 images (lower panel). Naturally, the ratio image is sensitive to the signal-to-noise ratio of the images divided. The ratio image is dominated by noise above and below the disk midplane, and at the inner and outer edges of the images. The structures seen in the inner disk ($<$2\farcs5) are dominated by PSF residuals which are different in the two images and result in noise. The ratio image, however, provides information on changes in the  disk properties between 2\farcs5 and $\sim$13". At these separations the ratio is close to 1.00, with variations typically below 3\%. The overall morphology of the ratio image does not suggest that any structure within the disk midplane has changed beyond these levels over our 15 year baseline.

\begin{figure}
\epsscale{0.95}
\plotone{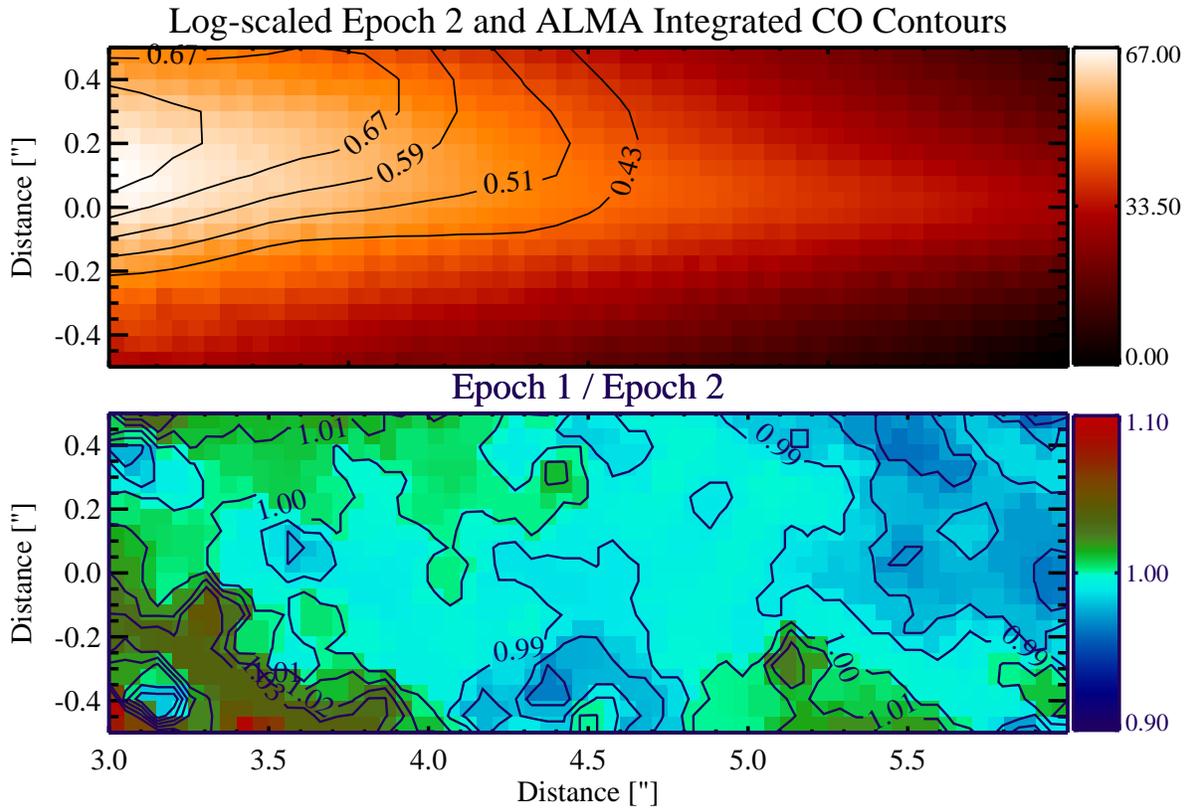}
\caption{{\em Top:} STIS direct image of the disk in logarithmic scale and velocity-integrated CO intensity from the ALMA observations by \citet[][]{Dent2014}. {\em Bottom:} Relative change between Epoch 1 and 2 in the SW wing. In the disk mid plane between 3--6" our measurements do not show changes greater than 3\%.  {\change The images have been rotated by 60.9$^\circ$ counter-clockwise from the standard north-up orientation to align the disk with the x-axis. }\label{Fig-Difference_Zoom}}
\end{figure}

Figure~\ref{Fig-Difference_Zoom} shows a magnified image focusing on the 3\farcs0--6\farcs0 region southwest from the star, extracted from the larger ratio image shown in Fig.~\ref{Fig-Difference}. This region covers most of the SW warp, the CO clump reported in \citet[][]{Dent2014}. The upper panel of this figure also shows the velocity-integrated ALMA CO emission overlaid in contours.  In order to improve the signal-to-noise ratio we binned the image by 2$\times$2~pixels using a flux-conserving IDL algorithm (frebin). 
The left edge of the image ($\sim3\farcs0$) is still influenced by the PSF residuals;  however, the rest of the image provides a higher quality measurement of the disk brightness evolution. This entire section of the image is within $\pm$2\% of 1.00, the level corresponding to no change. Given the small uncertainties in our photometric calibration we conclude that there is no evidence for changes in the disk surface brightness between 1997 and 2012 at levels 3\% or higher at the disk midplane or above at the location of the warp, the MIR clump, or the CO/sub-mm continuum clumps.


\begin{figure}
\epsscale{0.95}
\plotone{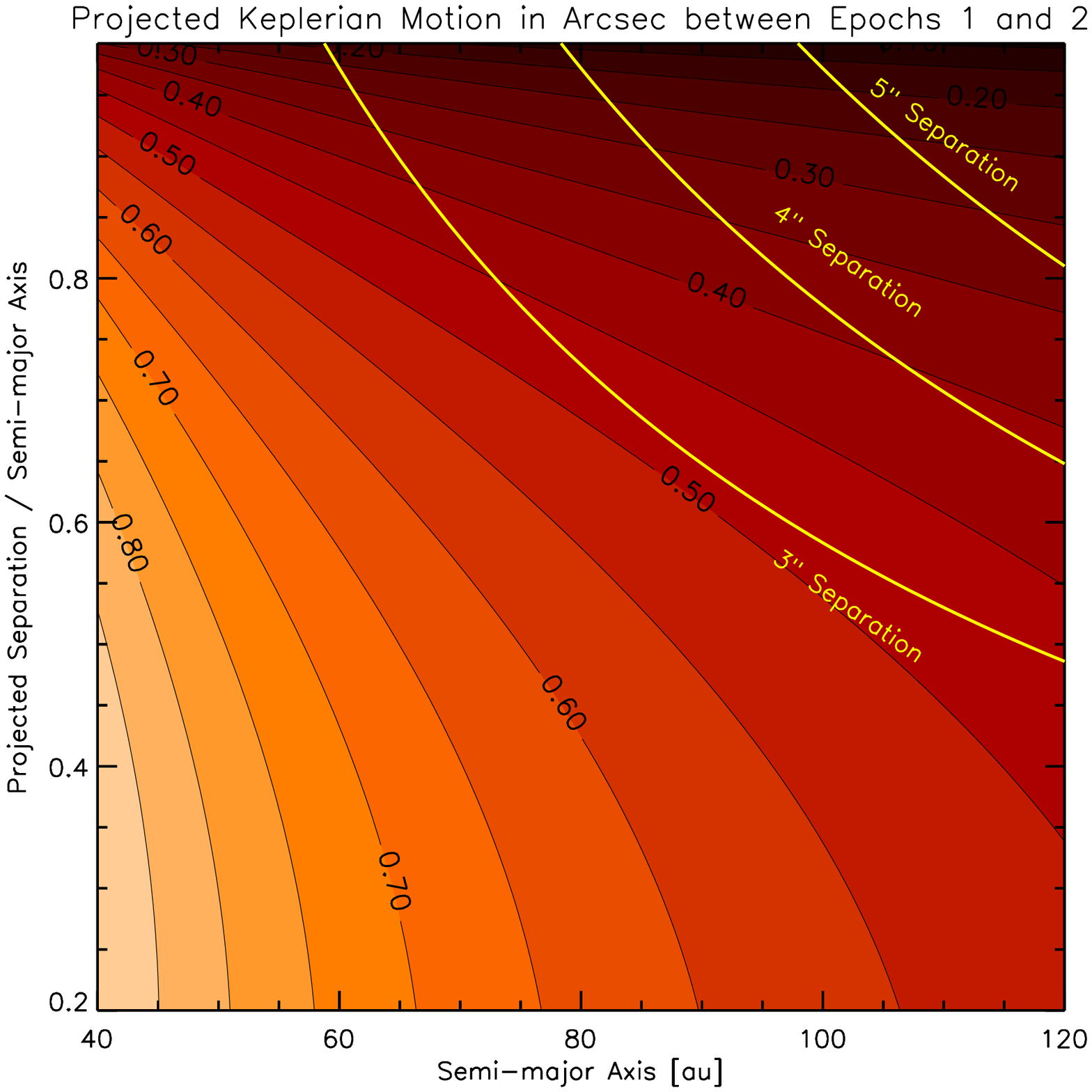}
\caption{The red-shaded contours show the difference in the projected location of a particle (in arc seconds) at a Keplerian orbit around \bpic between our Epoch 1 and 2 observations. The yellow curves show the curves corresponding to 3", 4", and 5" projected separations from \bpic. The STIS pixel scale is 0\farcs0507 and the predicted projected orbital motion is greater than a resolution element over the entire plot. Our observations place a 3\% upper limit at the difference in the mid-plane disk surface brightness over 15 years, effectively excluding the presence of single dust clumps contributing at this level to disk above the 3" yellow curve.     \label{OrbitalMotionTimescale}}
\end{figure}

In Fig.~\ref{OrbitalMotionTimescale} we compare the projected Keplerian motion for circular orbits over our STIS baseline (14.5 years) as a function of the orbital semi-major axis (x-axis) and the ratio of the projected separation and the semi-major axis (y-axis). For example, the upper edge of the plot (y=1) shows the projected motions if the planet is seen at quadrature; smaller y values correspond to narrower observer-star-orbital position angles. The red-shaded contours show the projected motion in arc seconds, ranging from 0\farcs1 to 0\farcs9 in the figure.  Yellow contours mark the x/y combinations that correspond to 3\farcs, 4\farcs0, and 5\farcs0 separations from the star. Based on the comparison of our Epoch 1 and Epoch 2 STIS images we excluded changes in the disk mid plane ($\pm$0\farcs25) greater than 3\% at separations 3\farcs0 to 5\farcs0.
The resolution elements of the STIS image is approximately 2 pixels or 0\farcs101. This value is smaller than the {\em smallest} projected Keplerian motion possible within the bounds of Fig.~\ref{OrbitalMotionTimescale}. 

Therefore, we conclude that any point source on a Keplerian orbit in or near the disk mid-plane contributing more than 3\% to the surface brightness distribution to the disk surface brightness, with projected separations between 3\farcs0 and 5\farcs0, would have led to a detectable change in our two-epoch comparison. This area covers part of the disk warp as well as the SW extension of the dust continuum seen in the ALMA images. It does not directly probe, however, the peak of the ALMA-detected integrated CO clump and the MIR-detected dust clump located at 2\farcs7 projected separation.

\subsection{Temporal Evolution Constraints on Extended Disk Structures}

{\change In this section we use a simple approach to translate the upper limits on change in disk surface brightness (see \S,\ref{S:TemporalChanges} and Fig.~\ref{Fig-Difference}) to upper limits on the time variations in the disk density structure considering a complicating factor, the disk's edge-on orientation. We synthesized some images of a generic dust cloud around \bpic assuming an isotropic scattering phase function. The model images contain azimuthal structures in the shape of simple cosine perturbations on the dust density, expressed in wavenumbers ($m$).  These structures rotate and shear with the Keplerian shear.  We compared these models  the region from 3--6" from the star on the SW side where our data is the best. 

The model has an unperturbed dust density with a radial power law index of $-1$, and a vertical distribution given by the \citet[][]{Kelsall1998} model of the solar zodiacal cloud. To simulate the perturbations, we multiplied the density distribution by the function:

\begin{equation}
1+ A \times cos(m \times (\Delta_{hel}-\Omega \times t) - \theta_{ripple})
\end{equation}

where $A$ is the amplitude of the ripple, $\Delta_{hel}$ is the stellocentric longitude, $m$ is the azimuthal wavenumber, $t$ is time, $\Omega$ is the local Keplerian angular frequency, and $\theta_{ripple}$ is the phase of the wave at time zero. This representation is equivalent to a cosine function that winds up with the Keplerian shear.

We created images at time $-7.5$ and $+7.5$ years later (via $t$, see Fig.~\ref{MarcKuchnerImages}) and took the ratio of the fluxes in a slice through the midplane. We then repeatedly rotated the disk slightly (via $\theta_{ripple}$), calculated the r.m.s. of the ratios to average over rotation angles.  We repeated this process for each of several angular wavenumbers and plotted slices to generate predictions for the signal as a function of radius. 

Only the $m=1$ wavenumber yielded any substantial signal in the 3--6"  region for amplitudes $A<1$.  Figure~\ref{MarcKuchnerImages} shows the rotation-averaged $m=1$ model compared to the ratio of \bpic images.  A dashed line shows the 3\% uncertainty in the ratio.  The $m=1$ model in this figure has an amplitude of 50\%, i.e. $A=0.5$.   This 50\% perturbation is marginally consistent with the detected change in the ratio.  An $m=1$ perturbation of this kind might be produced via an avalanche of dust from a single localized event -- a large collision, for example (see \citealt[][]{Kral2014}).

\begin{figure}
\epsscale{0.9}
\plotone{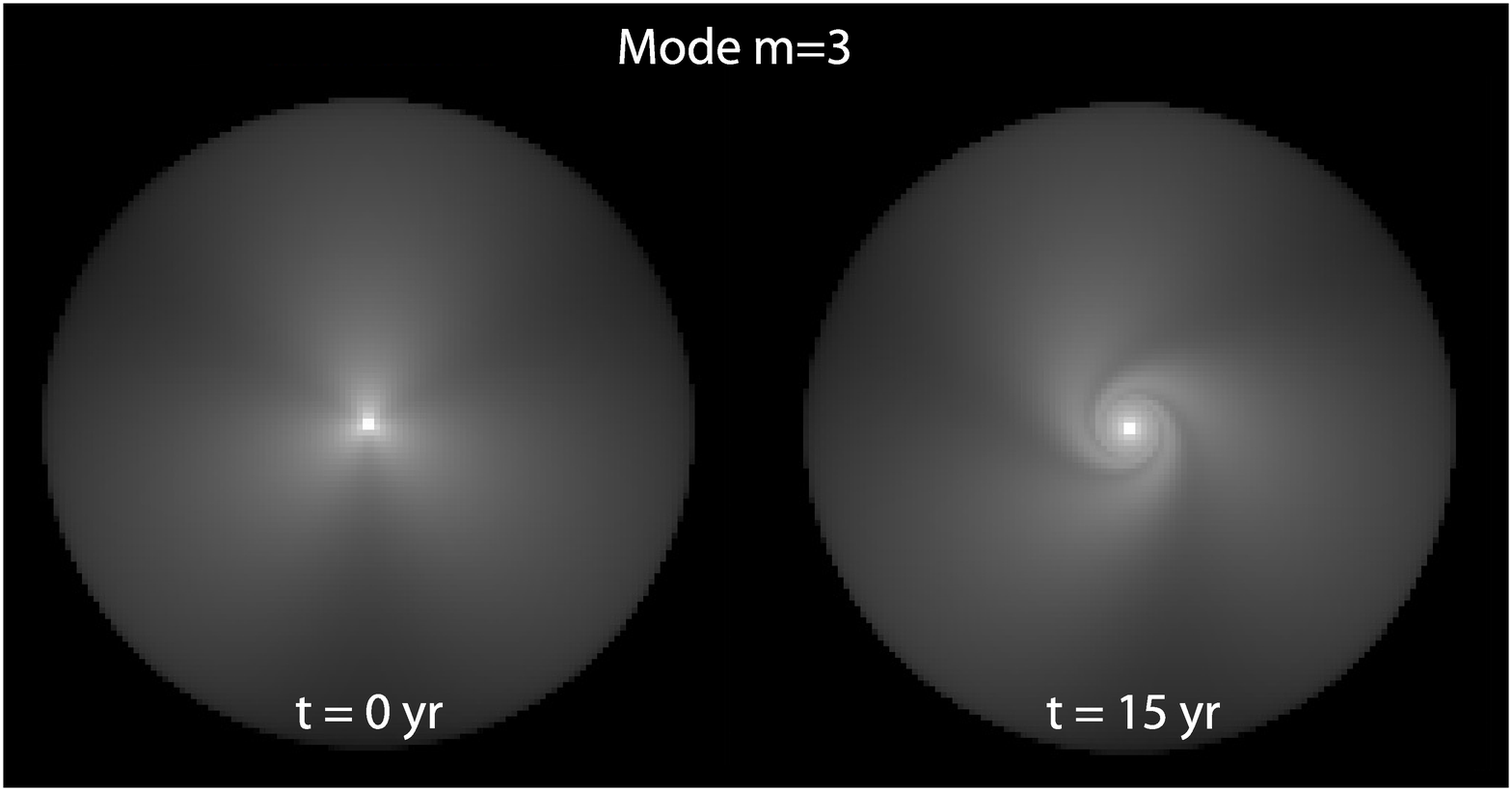}
\caption{Face-on images of the m=3 mode, at time 0 and 15 years later, in logarithmic scaling.
For the plot of the predicted temporal evolution signal shown in Fig.~\ref{MarcKuchnerPlot} edge-on images were used. \label{MarcKuchnerImages}}
\end{figure}

\begin{figure}
\epsscale{0.9}
\plotone{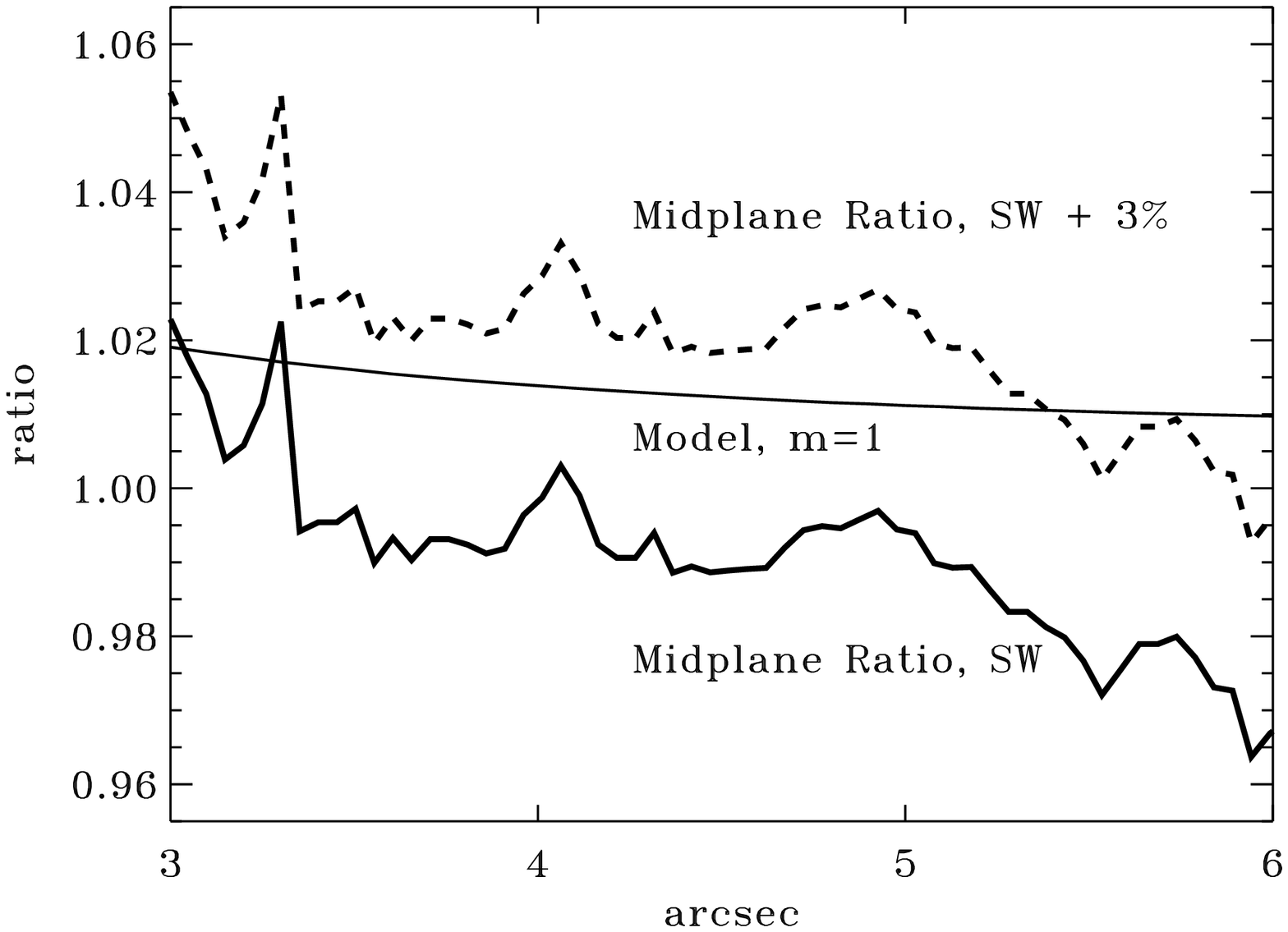}
\caption{Change in surface brightness between Epoch 1 and 2 for the angular wave number ($m=1$,
assuming 50\% perturbations, i.e. $A$=0.5). Our upper limit of 3\% beyond 3" argues for absent or very low-amplitude low-order wavenumbers.\label{MarcKuchnerPlot}}
\end{figure}

Our simple model shows that our tight upper limit on the changes ($<3\%$ at 3" and beyond) requires a disk that is devoid of high-amplitude, low-order angular modes and small, isolated structures; instead, \bpic appears azimuthally homogeneous at the radii probed.

We point out that the methodology described here can be used for future multi-epoch observations of disks and demonstrate that sensitive multi-epoch observations of nearby debris disks can provide powerful constraints on the azimuthal structure of the disk.

Our current observations are limited by the accuracy of the lower-quality first-epoch STIS image (1997). For the well-sampled disk regions (mid-plane, beyond 3\farcs0) our comparison yields an upper limit of only 2--3\%, which is the most accurate such comparison yet. Interestingly, even at these large separations and long periods the 14.5-year baseline results in large and easily-detectable projected Keplerian motions (see Fig.~\ref{OrbitalMotionTimescale}). The precision reached shows {\change that further} follow-up observations separated by just 3 years from our 2012 images would yield images capable of probing orbital motions at 2\farcs7, the location of the MIR- and CO-clumps.
Similarly, image pairs separated by only 3 years, taken by the James Webb Space Telescope or the ALMA sub-millimeter interferometer, should be able to sensitively probe Keplerian motions in the disk.
}

\section{Possible Origins of The Disk Structure in the $\beta$ Pic System}
\label{Sect:Models}
We now contrast disk structure models with the high-quality multi-wavelength images (\S \ref{S:DiskStructure}) that cover the \bpic disk and with the
improved orbital fits to \bpic~b (\S \ref{S:Orbit}). We also use simple models to interpret our upper limits on temporal variations (\S \ref{S:TemporalChanges}) and
demonstrate the future potential of such measurements.

\subsection{Planet-induced Disk Structures}

Multiple models have been proposed to explain the asymmetries of the \bpic disk. \citet[][]{Mouillet1997} and \citet[][]{Augereau2001} introduced a dynamical model that accounts for many of the observed properties of the \bpic disk and successfully predicted the presence of a massive planetary companion in the disk. This model assumed a giant planet at an orbit inclined by 3$^\circ$ with respect to a planetesimal belt, which extends up to $\sim$120~au. A collisional cascade within this planetesimal belt replenishes the dust grain population and the smallest grains (with sizes below the blow-out size) are efficiently moved  by radiation pressure from the planetesimal belt to longer orbits. These small grains are responsible for the very large scattered-light disk seen beyond $\sim$120~au.

In this model the warp in the inner disk is a direct result of the gravitational perturbation of the planetesimal belt by the giant planet. The inclined orbit of the giant planet forces the precession of the planetesimals' orbit, which leads to a warp. The outer radius of the warp will slowly grow in time, as the effect of the perturbations gradually accumulates for planetesimals on longer orbits. In addition, their simulations show the small dust grains generated by a collisional cascade in the warp are then subsequently blown out and will provide an extended, asymmetric structure, fully consistent with the prominent butterfly asymmetry.

The disk structure predicted by the  \citet[][]{Augereau2001} model continues to show an excellent agreement with the {\em axisymmetric} disk structures in \bpic, including the radial surface brightness profile (\S\, \ref{DiskSurfaceBrightnessProfiles}), the radially-normalized vertical disk profile (\S\, \ref{Section:VerticalStructure}) and the disk warp (\S\, \ref{SectionWarp}), as well as the large-scale structure and the inner disk structure (\S\, \ref{S:Orbit}). The model has only two assumptions: a planetesimal belt extending out to 120--150~au and the presence of a giant planet on an inclined orbit. 
The latter prediction has been verified through the detection of \bpic b and the former is supported by the submm observations (SMA: \citealt[][]{Wilner2011}, ALMA: \citealt[][]{Dent2014}).

\subsection{A Simple Model for the Warp Morphology}

The inner disk morphology offers important insights into the disk structure and we employ here a simple model aiming to explain the variation in the deviations above and below a linear disk mid-plane with stellocentric distance characterized by the brightest points along the disk major axis (see Fig.~\ref{VertExpansion}).
We explore the morphology of the warp in the inner disk by applying a model based on \citet[][]{Wyatt2005} and its extension (see, \citealt[][]{MatthewsPPVI}). We assume an edge-on planetesimal disk with the planetesimals' semi-major axes logarithmically spaced from 40--140~au, and assume that the disk surface brightness is scattered light that is directly proportional to the surface density of planetesimals. We placed a mass representing \bpic b on an inclined orbit ($i_{pl}=2^\circ$) and allowed the system to evolve for 10~Myr. We assumed that the disk surface brightness is scattered light that is directly proportional to the surface density of planetesimals whose mutual collisions produce the light-scattering dust. This is, of course, a toy model with several important simplifications: First, the system parameters (semi-major axes, inclinations, etc.) have not been optimized in any way; second, the model only follows planetesimals and not dust grains, which are affected by radiation pressure. Nevertheless, we expect the model to qualitatively reproduce the first-order morphology of the disk and thus serves a useful way to probe the \bpic disk morphology.
 
\begin{figure}
\epsscale{1.0}
\plotone{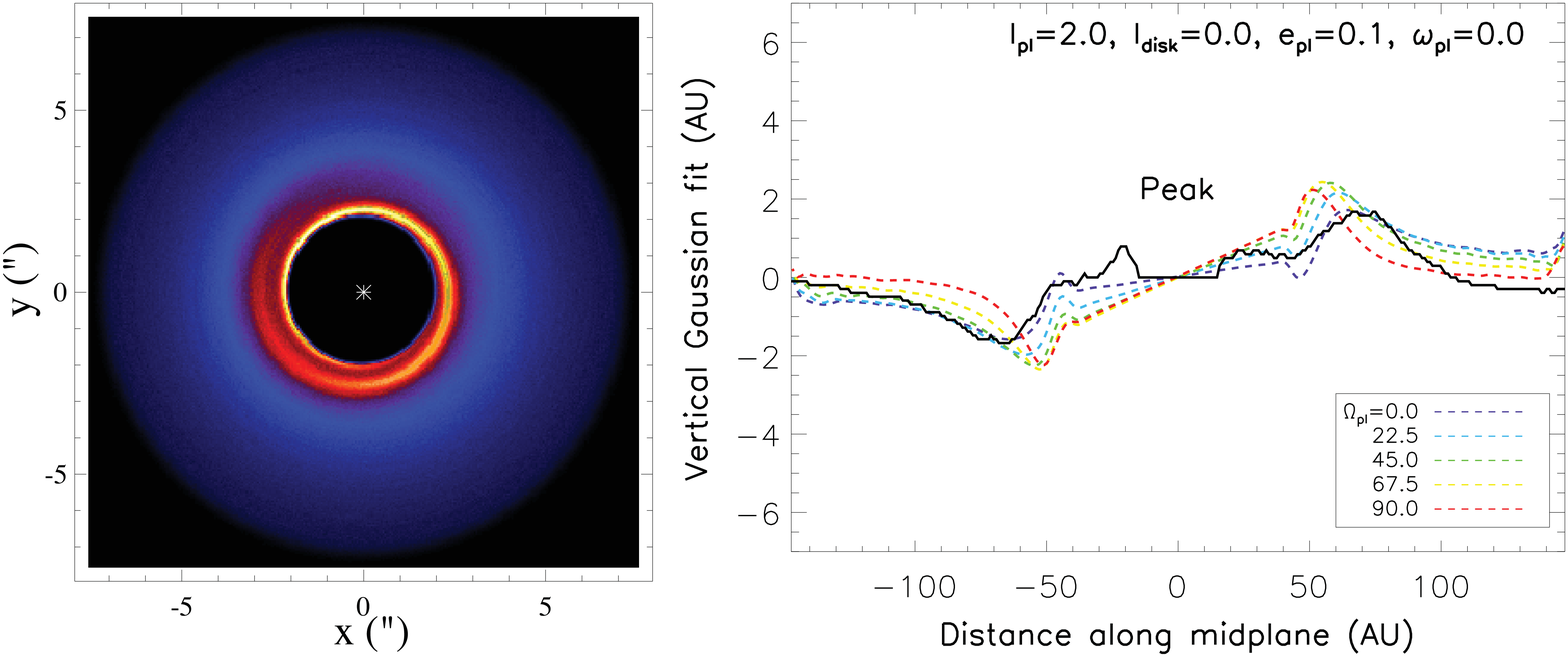}
\caption{ {\em Left:} Our simple model for disk-planet interactions predict a tightly wound spiral in the density distribution of the disk if the planet is on an eccentric orbit.  {\em Right:} The vertical location of the peak disk brightness at each radius as a function of the orientation of the line-of-nodes (in color) and the same observed quantity (thick solid line). \label{MarkWyattModels}}
\end{figure}

In the right panel of Fig.~\ref{MarkWyattModels} we show the vertical location of the peak brightness at each radius (peak brightness profile) from this model. We vary the line-of-nodes of the planet ($\Omega_{pl}$) and show the resulting peak brightness profiles compared to a  peak brightness profile derived from the observed disk image in an identical way. We note that this representation is very similar, but not identical to the more detailed fitting procedure described in \S\,\ref{Section:VerticalStructure}.

Our simple toy model leads to three conclusions: First, we find that the orbit and mass of \bpic ~b and the age of the system are such that the secular perturbations from the planet would be expected to impose a warp on the disk at a radial location and magnitude that is compatible with that observed. This conclusion is in line with previous studies \citep[e.g.][]{Augereau2001} and confirms the trend in spite of the simplicity of our approach. 

Second, our model demonstrates the importance of constraining the shape of the warp in the inner region explored by our STIS observations, because that shape constrains the orientation of the planet's orbital plane, specifically whether the line-of-nodes is in the plane of the sky ($\Omega=0^\circ$) or oriented along the line-of-sight ($\Omega=90^\circ$). If $\Omega_{pl}$=90$^\circ$ then the plane of symmetry in the inner disk would be a straight line between the warp. However, if $\Omega_{pl}$=0$^\circ$ then the inner disk would be aligned with the outer disk. 
Early models had noted that this parameter affected the magnitude of the warp \citep[][]{Mouillet1997}, but since that magnitude was maximized for $\Omega=90^\circ$ (Fig.~6 in that paper), later models only considered the $\Omega=90^\circ$ case. The fact that the Gaussian fits in Fig.~\ref{VertExpansion} find the inner regions to be well aligned with the outer disk seems to favor a model with $\Omega=0^\circ$ (see Fig.~\ref{MarkWyattModels}). 
{\change However, the orbital fits in Table~\ref{T:Orbitalelements} favour a value closer to $90^\circ$.}
To explore this further would require a detailed model which addresses the simplifications mentioned in the first paragraph of this section, which is beyond the scope of the current paper.

Third, our model also highlights the point that if the planet has an eccentric orbit then the same secular perturbations that result in the warp will also have caused the disk to have a tightly wound spiral structure (left panel in Fig.~\ref{MarkWyattModels}, e.g., \citealt[][]{Wyatt2005}, and also in \S\,4.3 of \citealt[][]{Mouillet1997}). The spiral does not significantly affect the structure of the warp because the evolution of eccentricities and inclinations are decoupled for low $e$ and $I$. However, the spiral could contribute a small brightness asymmetry to the disk. We note, that this spiral would be at the same radial location as the warp, since secular perturbations make orbital planes and pericenters precess at the same rate, which is also the same location as the clump. {\change A more comprehensive modeling of the time-evolving morphology introduced by the projection of the spiral structure would be valuable, but beyond the scope of the current paper.}

With the rapidly improving orbital period and mass estimate for \bpic and high-quality multi-wavelength images at hand, a more exhaustive study of \bpic disk dynamics is well motivated, but it is beyond the scope of the current paper.

\subsection{Open Questions: NE-SW Asymmetry and the Origin of a Super-jupiter on Inclined Orbit}

While all the axisymmetric structures observed in the disk are consistent with the predictions of the models describing the dynamical interactions of a planetesimal belt and a giant planet on an inclined orbit, these models cannot reproduce the prominent NE--SW asymmetries 
(see Table~\ref{T:Asymmetries} and \S\,\ref{S:DiskStructure}). This fact highlights the two fundamental open questions on \bpic: {\em What is the origin of the NE-SW asymmetries?} and  {\em What is the origin of the \bpic b super-jupiter's inclined orbit?} 
While our observations do not directly help to answer the latter question, they provide some constraints on the different scenarios proposed to explain the former. 

The mid-infrared observations by \citet[][]{Telesco2005} and \citet[][]{Li2012} show that the dust grain population in the bright dust clump in the SW wing is different (possibly smaller grains) than those typical to the rest of the disk. These authors argued for the possibility of a single, recent collision injecting copious amounts of dust in the system. The recent ALMA images by \citet[][]{Dent2014} show that the SW clump not only has a different dust population, but also contains gas-phase CO. Given that the UV--photodissociation lifetime of gas-phase CO by the interstellar UV field in an optically thin disk is $\sim$120 years \citet[][]{Visser2009}, the observed CO gas must have been released recently, consistent with a recent major collision (\citealt[][]{Telesco2005} and \citealt[][]{Dent2014}). The estimated mass in the SW dust clump detected by ALMA argues for a Mars-sized parent body, assuming that about 10\% of its mass has been released as debris \citep[][]{Dent2014}.  
A similar recent major collision has been proposed by \citet[][]{Stark2014} as one of two possibilities to explain the asymmetries observed in the HD~181327 debris disk.  To avoid the improbability of having witnessed such a massive collision only 120 years ago in \bpic, \citet[][]{Jackson2014} pointed out that the asymmetric dust distribution resulting from such an event can last up to 0.5~Myr if the dust clumps mark the location where the collision occurred, since debris released by the collision will have a variety of orbits but the orbits will converge at the collision point. However, this scenario would predict  that the dust clump's location is constant (i.e. not on Keplerian orbit), which is marginally inconsistent with the tentative detection of orbital motion of the clump \citet[][]{Li2012}.

An alternative scenario has been proposed for \bpic based on work by \citet[][]{Wyatt2003} and \citet[][]{Wyatt2006}. In this picture a planet -- during its outward migration -- captures planetesimals in mean motion resonances. The frequent mutual collisions of these planetesimals give rise to dust clumps that trace the planetesimal population and thus produce  resonant structures that co-rotate with the planet. This scenario predicts a $>$10~M$_{Earth}$ planet at the inner edge of the dust belt (with half the orbital period as that of the belt). This scenario has been proposed for \bpic both by \citet[][]{Telesco2005} and \citet[][]{Dent2014}.

Perhaps the most directly testable key difference between these two scenarios is the SW dust clump's orbital motion or its absence: if the clump is on orbit (as suggested by \citealt[][]{Li2012}) then this would argue for it being a resonant structure. If its location is constant, the clump would likely be a result of a recent collision. 

It is important to point out that neither of the above scenarios for the origin of the SW emission invoke the dynamics proposed to explain the warp. Yet, the warp is at exactly the same radial location as the clump, which may be either a coincidence or there is a causal connection. There may be scenarios in which the warp aids the formation of the clump. For example, if \bpic~b is on an eccentric orbit then there will be a spiral pattern at the location of the warp. The resulting brightness asymmetry is much smaller than that observed. However, we also know that this is also the location where orbits of debris that were previously not overlapping start to cross \citep[][]{MustillWyatt2009}, which might increase the chances of seeing a recent collision at that location. For the second scenario there is not a causal connection, but the spatial coincidence is not surprising, since the orbit of the outer planet (that with the resonantly trapped planetesimals) would be aligned with \bpic~b and would also have aligned the resonant planetesimals with its orbit.


\section{Summary}
\label{Summary}
We present new and re-reduced archival HST/STIS coronagraphic optical imaging of the $\beta$~Pictoris debris disk. Our data provide the yet highest signal-to-noise and smallest inner working angle optical images of the disk. Based on our images we characterize in detail the
disk structure and compare the optical images to published images taken at wavelengths ranging from the near-infrared to mm.

The key findings of the paper are the following:

1) Our high-contrast images provide a continuous coverage of the inner disk covering radii from 0\farcs35 to 13" in unfiltered light. The inner working angle is about $2\times$ smaller than that of previous optical images and allows us to image the disk at a location and time where the \bpic b planet was present in 2012. 

2) We identify radial regions with constant radial surface brightness slopes. We show that corresponding
regions in the northeastern and southwestern wings have similar, but slightly different surface brightness slopes, indicating an asymmetry in the disk.

3) We find that the surface brightness slope between 0\farcs4 -- 2\farcs0 is constant, arguing against any significant changes in the disk structure at or adjacent to the most likely semi-major axis of \bpic b. 

4) We present the first optical images of the inner structure of the main disk and its vertical extension. {\change While our images do not show the presence of a separate secondary disk as suggested by \citet[][]{Golimowski2006}, preliminary comparison to two-component scattered light disk models suggest that our observations are consistent with the warp caused by the projection of an inner, inclined disk onto the outer disk.}

5) NE--SW asymmetry: we confirm that the two wings of the disk have different radial surface brightness slopes and that the SW wing is brighter in the inner {\change 8\farcs0} but fainter beyond that than the NE wing. We show that the same asymmetry extends to the innermost disk, down to at least 0\farcs5.

{\change 6) The angle of the disk warp (seen in projection) is approx. $4^\circ$, significantly larger than the best-fit inclination of \bpic b's orbit ($0.7^\circ$). }

7) Careful comparison of STIS images obtained over 15 year baseline (in 1997 and in 2012) shows no difference greater than 3\% in the disk surface brightness at the locations of the disk warp part of the CO and sub-mm continuum clumps (SW 3--5"). Similarly, we do not detect any difference between the radial surface brightness profiles.

8)  We compile all disk asymmetries seen in wavelengths ranging from optical through mid-infrared to sub-mm wavelengths. We divide the asymmetries in two groups: apparently axisymmetric and apparently non-axisymmetric. The axisymmetric disk structures appear to be fully consistent with the structures predicted by models of \citet[][]{Mouillet1997} and \citet[][]{Augereau2001}, but the axisymmetric structures argue for either a recent major collision or the presence of plantesimals on resonant orbits with a yet unseen planet at $\sim$80~au.

9) We show that over a few year baseline projected Keplerian motions are sufficiently large to allow detection with existing and near-future facilities. We argue that, in particular, HST, JWST, and ALMA multi-epoch observations will provide powerful constraints on the azimuthal structure, dynamics, and disk-planet interactions in the \bpic and other nearby debris disks.

Our work leads to two fundamental questions required to explain the properties and origin of the \bpic system: {\em How did super-jupiter \bpic b ended up on an inclined orbit?} and {\em What is the origin of the NE--SW asymmetry?} We discuss different scenarios and show that time-resolved studies of Keplerian motions in the \bpic can be a powerful way to discriminate these.

\acknowledgments
We thank STScI program coordinator Tricia Royle, contact scientist John Debes, and Charles Proffitt for their dedicated support of this program. {\change We thank Jean-Charles Augereau, Rebecca Dawson, Andras Gaspar, Paul Kalas, John Debes, Bill Dent, and Kate Su, among others, for valuable discussions. We thank the anonymous referee, whose timely report has helped to improve the interpretation of our results and the clarity of the manuscript.}
We thank the entire Servicing Mission 4 crew for restoring HST and STIS operations.
{\change We are grateful to the anonymous referee for a prompt and detailed review which helped to clarify the paper. The paper also benefitted from the presentations and discussions at the workshop Beta Pictoris at 30, Paris. }
Support for Program number 12551 was provided by NASA through a grant from the Space Telescope Science Institute, which is operated by the Association of Universities for Research in Astronomy, Incorporated, under NASA contract NAS5-26555. MCW is grateful for support from the EU through ERC grant number 279973.We acknowledge support from the French National Research Agency (ANR) through the grant ANR10-BLANC0504-01.

{\it Facilities:} \facility{HST (STIS), \facility{ALMA}, \facility{VLT (NACO}, \facility{Gemini (TRECS}}.

\clearpage

\bibliographystyle{aa}       
\bibliography{apairefs}   

\end{document}